\documentclass[a4paper]{article}

\usepackage[pages=all, color=black, position={current page.south}, placement=bottom, scale=1, opacity=1, vshift=5mm]{background}

\usepackage[margin=1in]{geometry} 

\usepackage{amsmath}
\usepackage{amsfonts}
\usepackage{lipsum}
\usepackage{algorithm2e}
\usepackage{graphicx}
\usepackage{subcaption}
\usepackage{float}
\usepackage{pifont}
\usepackage{bbding}
\usepackage{epstopdf}
\usepackage[export]{adjustbox}

\usepackage{subcaption}

\usepackage[utf8]{inputenc}
\usepackage{hyperref}
\hypersetup{
	unicode,
	pdfauthor={Vahid Tavakol Aghaei},
	pdftitle={Energy Optimization of Wind Turbines via a Neural Control Policy Based on Reinforcement Learning Markov Chain Monte Carlo Algorithm},
	pdfsubject={A simple article template},
	pdfkeywords={Reinforcement Learning, Energy Maximization, Wind Turbine, Control, Markov chain Monte Carlo},
	pdfproducer={LaTeX},
	pdfcreator={pdflatex}
}

\usepackage[authoryear,round]{natbib}

\usepackage{graphicx, color}
\graphicspath{{figures/}}

\usepackage{algorithm, algpseudocode} 
\usepackage{mathrsfs} 

\usepackage{lipsum}

\title{Energy Optimization of Wind Turbines via a Neural Control Policy Based on Reinforcement Learning Markov Chain Monte Carlo Algorithm}
\author{Vahid Tavakol Aghaei$^{1,2}$\thanks{Corresponding author} \and Arda Ağababaoğlu$^3$ \and Biram Bawo$^4$ \and Peiman Naseradinmousavi$^5$ \and Sinan Yıldırım$^6$ \and Serhat Yeşilyurt$^7$ \and Ahmet Onat$^8$}

\date{
	$^{1*}$Istinye University, Faculty of Engineering and Natural Sciences, Electrical and Electronics Engineering, Istanbul, Türkiye\\ \texttt{vahid.aghaei@istinye.edu.tr}\\[2ex]%
	$^2$Center of Excellence for Functional Surfaces and Interfaces for Nano-Diagnostics (EFSUN), Sabanci University, Orhanli, 34956, Tuzla, Istanbul, Türkiye\\[2ex]
	$^3$Delivers AI, İTÜ Arı Technological Park, Istanbul, Türkiye\\ \texttt{agababaoglu@sabanciuniv.edu}\\[2ex]%
	$^4$Sabanci University, Faculty of Engineering and Natural Sciences, Mechatronics Engineering, Istanbul, Türkiye\\ \texttt{bawobiram@sabanciuniv.edu}\\[2ex]
	$^5$San Diego State University, Dynamic Systems and Control Laboratory, San Diego, USA\\
	\texttt{pnaseradinmousavi@sdsu.edu}\\[2ex]%
	$^6$Sabanci University, Faculty of Engineering and Natural Sciences, Industrial Engineering, Istanbul, Türkiye\\
	\texttt{sinanyildirim@sabanciuniv.edu}\\[2ex]%
	$^7$Sabanci University, Faculty of Engineering and Natural Sciences, Mechatronics Engineering, Istanbul, Türkiye\\
	\texttt{syesilyurt@sabanciuniv.edu}\\[2ex]%
	$^8$Istanbul Technical University, Control and Automation Engineering, Istanbul, Türkiye\\
	\texttt{ahmetonat@itu.edu.tr}
}

\begin{document}
	\maketitle
	
\begin{abstract} 
This study focuses on the numerical analysis and optimal control of vertical-axis wind turbines (VAWT) using Bayesian reinforcement learning (RL). We specifically address small-scale wind turbines, which are well-suited to local and compact production of electrical energy on a small scale, such as urban and rural infrastructure installations. Existing literature concentrates on large scale wind turbines which run in unobstructed, mostly constant wind profiles. However urban installations generally must cope with rapidly changing wind patterns. To bridge this gap, we formulate and implement an RL strategy using the Markov chain Monte Carlo (MCMC) algorithm to optimize the long-term energy output of a wind turbine.
Our MCMC-based RL algorithm is a model-free and gradient-free algorithm, in which the designer does not have to know the precise dynamics of the plant and its uncertainties. Our method addresses the uncertainties by using a multiplicative reward structure, in contrast with additive reward used in conventional RL approaches.
We have shown numerically that the method specifically overcomes the shortcomings typically associated with conventional solutions, including, but not limited to, component aging, modeling errors, and inaccuracies in the estimation of wind speed patterns. Our results show that the proposed method is especially successful in capturing power from wind transients; by modulating the generator load and hence the rotor torque load, so that the rotor tip speed quickly reaches the optimum value for the anticipated wind speed. This ratio of rotor tip speed to wind speed is known to be critical in wind power applications. The wind to load energy efficiency of the proposed method was shown to be superior to two other methods; the classical maximum power point tracking method and a generator controlled by deep deterministic policy gradient (DDPG) method.
\end{abstract}

\maketitle
\section{Introduction}
The main purpose of machine learning (ML) is to 
determine action policies to perform successfully in unknown environments, or tune control strategies to optimize a design goal.
One of the most effective ML methodologies, also used for solving control problems, is reinforcement learning (RL) \citep{Sutton_and_Barto_1998}. In RL setting, an agent (controller) applies an action to 
achieve state transitions after which an immediate reward portraying the performance of the agent is issued. 
An optimal policy of selecting actions that optimize the cumulative reward is then sought.
Recently,
renewable energy sources have received a lot of attention as a result of increasing demands towards the ecological concerns, where wind energy conversion technologies have piqued much interest 
among 
the various renewable energy sources \citep{CHENG2014}.


The most prevalent forms of wind turbines are horizontal and vertical axis ones (HAWT and VAWT) in which the rotor's axis is either 
oriented 
horizontally or vertically with respect to the wind stream 
have been 
thoroughly investigated by \cite{aykut} in terms of their merits.
	In parallel 
	to the wide-spread increase in the applications of wind energy in every possible sector, the main focus of this research is 
	the numerical analysis and optimal control of VAWTs 
	where their operational condition is independent of wind direction. Furthermore, since they produce less noise and need less maintenance, 
	they can be well-suited to urban and remote 
	power generation areas, 
	especially in small-scale 
	applications \citep{KHORSAND2015,TUMMALA20161351,TASNEEM2020}. 
	Through this work, we formulate and implement a RL strategy using Markov Chain Monte Carlo (MCMC) algorithm to optimize the long-term energy output of the wind turbine. The method specifically overcomes the shortcomings typically associated with conventional solutions including but not limited to component aging, modeling errors and inaccuracies in the estimation of wind speed patterns. Our RL-MCMC algorithm is a model-free and gradient-free algorithm, where the designer does not have to know the precise dynamics of the plant and their uncertainties. The method has been observed to be especially successful in capturing power from wind transients; it modulates the generator load and hence rotor torque load so that the rotor tip speed reaches the optimum value for the anticipated wind speed. This ratio of rotor tip speed to wind speed is known to be critical in wind power applications.
Recently, thanks to advancements in electronic and power control devices,  variable-speed control \citep{DALI20211021} for wind energy conversion systems (WECSs) has enabled greater energy harvesting from the wind. Pitch angle of the rotor as well as electrical load are commonly used to regulate the speed of a WECS. Various variable-speed control algorithms such as sliding mode control (SMC) \citep{YANG2018SMC}, MPPT \citep{HU2019,SITHARTHAN2020}, model predictive control (MPC) \citep{onol2015model11}, adaptive neuro-fuzzy \citep{ASGHAR2018}, and RL \citep{7370922RL} have been implemented on wind turbines.



MPPT is a famous control method for VAWTs. Each turbine running at a certain wind speed has an optimal tip-speed ratio (TSR) that corresponds to a specific generator rotor speed ($\omega_{r}$) and yields maximum power. It is this ratio and its derivative that MPPT algorithms strive to optimize. While MPPT is effective at maximizing the instantaneous power, this is not the same as maximizing the whole energy available. The three main MPPT algorithms which are elaborately addressed in \cite{Lasheen}, are the TSR, the perturb and observe (PaO), and the power signal feedback control. Moreover, there are other MPPT methods in the literature such as hybrid-adaptive PaO \citep{YOUSSEF2020}, fuzzy logic based MPPT with a grey-wolf optimization algorithm \citep{LAXMAN2021GREYWOLFFLC, seyyedabbasi2021gwo}, and sensorless MPPT algorithms \citep{LI2019}. Specifically, to control the small-scale wind turbines, there exist fuzzy-based MPPT \citep{Ngo2020TheMP,YAAKOUBI2019}, PaO MPPT \citep{en1220} and limited power point tracking (LPPT) \citep{app102280} methods.



MPC is another typical control method utilized in the operation of VAWT. With a convex objective function and an accurate system model, this strategy can be helpful in achieving the desired results \citep{GARCIA1989335}. For obtaining a wind profile-optimized control signal, this method employs a finite-horizon prediction approach. Due to the fact that MPC addresses the optimization problem using a finite-horizon, there are several circumstances in which the objective function might not be optimized as it should be. Additionally, in order to comply with MPC regulations, the future data of wind speed must be gathered and individually transmitted to the wind turbine. In cases of dramatic wind variations, MPC may not be effective due to its sluggish performance. The research that has been done on MPC for WECS has resulted in multiple different applications \citep{9485106MPCWECS,SONG2017564}. Despite their success, the challenge with MPC algorithms is threefold; a time horizon for control predictions must be selected, a plant model that is accurate through that horizon is required, and high computational cost is incurred for evaluating the model \citep{Bemporad15}. Costs associated with MPC calculations are large, even with ML-based models. Moreover, real-time applications may not be possible depending on the control problem's convexity and complexity \citep{NOROUZI2022105299}.

\begin{figure}
	\centering
	\includegraphics[scale=0.16]{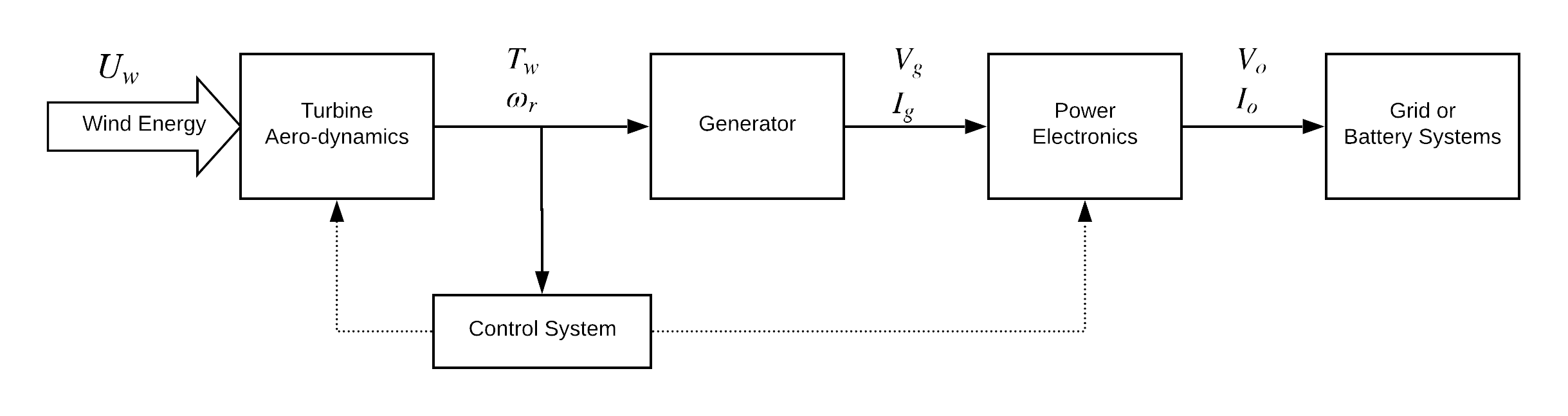}
	\caption{A Block representation of a   controlled WECS.}
	\label{fig:WECS}
\end{figure}




Achieving maximum control efficiency in modern control and robotic systems is a priority, hence researchers have recently turned to ML techniques \citep{OUYANG20221}. Some researches \citep{Bui2021RL,ZHANG2020113063,9457135RLLin,KOFINAS2017461} incorporate RL into MPPT to improve the efficiency of MPPT; this is noteworthy given the paucity of efforts dedicated to ML-based control of WECS. Furthermore, some other works, have used artificial neural-networks (ANN) to control WECS \citep{10.1007/978-981-16-7664-2_35, CHOJAA2021ANN}.
\cite{ANNSUN2020} used ANN models to predict the output power of a wind turbine based on recorded experimental data (wind direction, speed, and yaw angles). Compared to these methods, the MCMC-based RL algorithm can incorporate expert knowledge into the learning process through its prior distributions, which can facilitate both the learning and accuracy of the learned models. On the other hand, the RL-MCMC algorithm is more data-efficient than the ANN because it requires less data for successful performance. This is because it can learn from previous experiences by exploring the high-reward areas of the search space and adjusting the model accordingly.

To maximize the energy output in wind turbines there exist some optimal control methods \citep{6240758,PARK2015295}. These methods formulate an analytic power function for a specific wind turbine model. However, they have the drawback of being susceptible to unmodeled dynamics and system uncertainties, which means that in practice, the performance of these model-based controllers may differ significantly from the findings of analytical calculations.  Without having the explicit model of the system, model-free control attempts to accomplish control goals by relying solely on the input and output data. Therefore, it can perform difficult tasks that are challenging to tackle using model-based methods and have excellent adaptive abilities to the underlying dynamics of complex systems. Regardless of the explicit model of the system, model-free control attempts to accomplish control goals  by relying solely on the input and output data. Therefore, it can perform difficult tasks that are challenging to tackle using model-based methods and have excellent adaptive abilities to the underlying dynamics of complex systems. These realities make model-free methods for optimizing wind power generation attractive alternatives.
As an alternative, a data-driven algorithm is proposed in \cite{gradientMPPT} which is based on the gradient optimization of the MPPT method. Moreover, recently \cite{DONG2021DDPGwind} proposed a deep RL algorithm based on a policy gradient method for the problem of the wind farm control. However, gradient-based algorithms may suffer from getting trapped in local optima and the need for advanced gradient-free algorithms is inevitable as thoroughly investigated by \cite{qian2021derivativefree}.
\subsection{Motivation, Gaps and Contributions}
When it comes to finding the optimal policy in stochastic dynamical systems, Bayesian RL has proven to be an efficient solution, especially for the control and robotics domains \citep{pmlr-v115-derman20a,9361111BRL, DBLP:journals/corr/abs-2107-09822, TAVAKOLAGHAEI2021}. However, there is much room for improvement regarding its potential use in energy conversion systems which arouses our curiosity to contribute to filling this gap. Some of the available Bayesian methods use Gaussian processes (GP) to learn either the model of the system or the desired objective function via learning the hyper-parameters of the GPs \citep{BayesianGP2010,WilsonFT14}. In this regard \cite{BayesianWECS, PARK16Applied} recently applied a Bayesian method to obtain an approximate model of the target function using GPs to maximize the wind farm power. However, 
	our method differs in that it is a model-free RL-based algorithm, which uses MCMC as a sampling strategy to draw samples from an instrumental function called \emph{posterior distribution}. By using the \emph{Bayes' theorem} we construct the posterior distribution, which is proportional to the objective function of the RL algorithm and some \emph{prior} density functions of the unknown parameters of the radial basis function neural networks controller (RBFNN). It is shown that the drawn samples from this posterior function will ultimately converge to the exact target function \citep{Andrieu:2003}. One disadvantage of the proposed Bayesian algorithm by \cite{BayesianWECS, PARK16Applied} is the computational cost of the GP regression model, whereas this issue is tackled in our proposed RL-MCMC method since we do not model the cost function and instead we build a posterior function which we can draw samples from and is guaranteed to converge to the desired target function. Another disadvantage with their method is the problems related to the \emph{trust-region} optimization method which they impose some constraints over the parameter search space. This results in finding local optimal solutions that are close to their initial parameters. Unlike their method, and inspired by the \emph{exploration-exploitation} problem, we have developed an RL-MCMC algorithm that benefits from the strengths of both the gradient-free Bayesian MCMC and RL algorithms. We use MCMC sampling method that is capable of exploring the \emph{high-reward} regions of the policy parameter space (benefiting from the long-term rewards in RL). Since the policy space is explored in a non-contiguous manner, different regions can be visited and the probability of discovering better performing regions always exists. The main point is that, our proposed Algorithm \ref{alg: Pseudo-marginal Metropolis-Hastings for reinforcement learning} is not designed to converge to a single point. Instead, the policy parameters are guaranteed to follow the probability distribution $\pi(\theta)$ stated in Equation \eqref{eq: poli_dist}. In terms of being a \emph{gradient-free} algorithm, there already exists some algorithms to analytically evaluate a target function \citep{derivfree1,ChangHW13, Rios2013DerivativefreeOA}. However, in our case, the objective function given in Equation \eqref{eq: J theta} is not analytically known. As a result we propose to use another alternative as the MCMC sampling method. Moreover, for the above-mentioned algorithms, a \emph{trust-region} is defined as a constraint over the search space which may guide the optimization towards local optima. It should also be noted that our Bayesian RL-MCMC algorithm has the advantage of being independent of the aerodynamic model of the system, considering the fact that a model-based control has weaknesses about the fluctuations in wind turbine parameters given that the power production is dependent on a variety of factors \citep{SOLEIMANZADEH2011720}.
In the literature, there exist numerous statistical and intelligent methods that focus on parameter estimation of solar systems \citep{TERRENSERRANO2021pv,KUMARI2021LSTM,ELIZABETHMICHAEL2022}, wind speed  \citep{CHEN2014,wang2018multi,ZHANG2019wind,QIN2019262,LIU2020114259,BAI2021117461,YU2022119692,HAN2022118777,ZHANG2022117815} and power prediction \citep{YU2019windpow,MA2022windpow,WANG2022windpow}. However, we use Bayesian RL which benefits from MCMC sampling method to learn the unknown parameters of the nonlinear controller of a wind turbine system to maximize the available output energy and we believe that the application of RL based Bayesian learning using MCMC is still in its infancy and needs to be explored more. On the other hand, most of the existing methods only deal with the available data sets. However, for our case we collect the data trajectories using RL algorithm in a sequential manner. This gives flexibility to control the system at hand and obtain the desired performance that a customer can expect from the designer.
The optimization of the output energy of WECS is one of the most challenging issues to be addressed. Generally speaking, the classic control methods for wind turbine systems are not designed with long term optimization in-mind.
It is important to devise an online smart optimization strategy that will enable an efficient control mechanism to be applied in the case of fluctuating wind patterns. Additionally, it is important to note that not every algorithm is optimal for systems which are linearized. Therefore, standard wind conversion systems face persistent difficulties in adapting to changing wind conditions.

Our proposal is to use a hybrid artificial intelligence and ML-based methodology that account for nonlinearities and uncertainties to address these challenges. At the heart of the control policy, a nonlinear RBFNN controller is developed in which its unknown parameters are being learned by an RL-MCMC method proposed in \cite{doi:10.1080/21642583.2018.1528483}. In the current paper however, we reshaped that original method and improved it through the inclusion of RBFNN as a nonlinear controller that will enable us to learn to control the unknown WAVT system. Similar to our control structure, recently, \cite{KEIGHOBADI2022} used a SMC with an RBFNN to compensate for the uncertainties and noises involved in the control of the wind turbines. However for our case, RBFNN is used as a neural control policy and we incorporate the system uncertainties into the long-term reward function of the RL algorithm, $J(\theta)$. Based on a known load current and voltage ($I_{L}$, $V_{L}$), rotor speed $\omega_{r}$, wind speed $U_{w}$ and its derivative, the proposed algorithm is capable of learning to control the VAWT's unknown model and can obtain the immediate control effort $I_{L}$ similar to the optimal load coefficient $C_L$. It also facilitates recognizing different variations (as friction, tear and wear of the blades, and elements aging) in VAWT dynamics over time.

Our focus is on a small-scale VAWT with a 3-bladed rotor structure. Comparing MPPT with the proposed RL-MCMC algorithm, demonstrated the efficiency of the proposed method. Furthermore, to validate our methodology a comparison in terms of output power is made with a famous state-of-the-art deep RL algorithm namely, Deep Deterministic Policy Gradient (DDPG).

The contributions of the current paper, inspired by the need for novel configurations for the renewable energy systems capable of optimizing the output energy, can be summarized as follows:
\begin{enumerate}
	\item {\bf ML-based control of turbine and its responses to wind patterns}
	\begin{enumerate}
		\item To learn to control VAWT's unknown dynamics, the proposed control structure employs a nonlinear RBFNN controller (calculating the reference load current $I_{L_{ref}}$), which its unknown parameters are learned using RL-MCMC. 
		\item Provides a way of 
		continuously adapting the controller to the changing wind patterns
		which makes it possible to the designer to install the VAWT in any location and still obtain the maximum available energy from the wind.
	\end{enumerate} 
	\item {\bf Developing a Bayesian gradient-free and model-free algorithm}
	\begin{enumerate}
		\item Shape of the objective function  in Equation \eqref{eq: J theta} is not explicitly known. Therefore, gradient and Hessian approximations cannot be a suitable choice for analytical solutions. In the considered setting, function evaluations by various simulations are costly. Thus, by leveraging MCMC, we could guide the search mechanism to the regions where obtaining the optimal parameters are most likely (high-reward areas).
		\item Developing a model-free algorithm which operates regardless of the aerodynamics model of system.
	\end{enumerate}
	\item {\bf Comprehensive simulation studies and comparisons with MPPT algorithm}
	\begin{enumerate}
		\item Maximize total energy output rather than existing greedy algorithms which typically aim to maximize instantaneous output power.
		\item Based on three different simulation scenarios, we demonstrate that the proposed RL-MCMC algorithm with an RBFNN controller outperforms the well-known classical MPPT.
	\end{enumerate}
\end{enumerate}
In particular, our MCMC-based RL algorithm is a model-free and gradient-free algorithm, where the designer does not have to know the precise dynamics of the plant and their uncertainties. We formulate the overall effect of such uncertainties into an expected total reward $J(\theta)$, which is given in Equation \eqref{eq: J theta}. It maximizes $J(\theta)$ of an instantaneous reward function $r(t)$ created by the designer, and calculates an action policy $\pi_{\theta}$, where $\theta$ is the set of controller parameters. We can take, for example, the instantaneous electrical output power as part of the reward function and our RL-MCMC algorithm maximizes its cumulative value, e.g., total energy output. To explore policies according to $J(\theta)$, the algorithm does not have to evaluate $J(\theta)$ explicitly; but by merely approximating it via MCMC. This is in fact one of the advantages of our methodology which we stress throughout the manuscript.
\subsection{Outline}\label{susec: Outline}
The rest of the paper is organized as follows: Section \ref{sec: WAVT_Model}  discusses the structure of the studied VAWT model, explaining the device configuration (Three-phase and single-phase permanent magnet synchronous generator PMSG and load model), parameters and its mathematical model. In this section the relation between the tip-speed ratio and coefficient of power for the wind turbine is given. Section \ref{sec: RL and MCMC} gives insights for the RL and MCMC methodologies and their applications to the dynamical systems are discussed. In this section the proposed RL-MCMC algorithm is provided as summarized in Algorithm \ref{alg: Pseudo-marginal Metropolis-Hastings for reinforcement learning}. Section \ref{sec: proposed_control} describes the proposed control strategy that includes the configuration of the proposed RBFNN controller with its unknown standard deviation and weight parameters to be learned. To clearly convey the working principles of the proposed control strategy, a black diagram representation is depicted. Then this section is concluded with the training steps of the learning RL-MCMC algorithm. The simulation results and their discussion are divided into four main subsections gathered under Section \ref{sec: results} where subsection \ref{subsec: 1} gives the reward function and the parameters of the prior distribution of the RL-MCMC algorithm. In subsection \ref{subsec: train1}, the first training stage of the RL-MCMC with a step wind input to the VAWT is presented and the simulation results of the learned parameters of the RBFNN controller, generated mechanical power, load current, load voltage, rotor's angular velocity, and load resistance plots are provided. Similarly, in subsection \ref{subsec: train 2}, the aforementioned time-response plots are obtained using a sinusoidal wind speed. For both subsections, the simulations have performed in two scenarios in which first the initial parameters of the RBFNN are chosen randomly and then the RL-MCMC algorithm is trained. At the end of training, it ends up with the new learned parameters and then uses those parameters to calculate the final time-response plots. To show the effectiveness and validity of our proposed RL-MCMC algorithm, comparisons are made with the classical MPPT and a deep RL algorithm in subsection \ref{sec: mppt,DRL}. Finally, Section \ref{sec: conc} concludes the paper and provides some discussions regarding the current methodology and results and gives some future research directions.

\section{VAWT Model}\label{sec: WAVT_Model}
\begin{figure}
	\centering
	\includegraphics[scale=0.8]{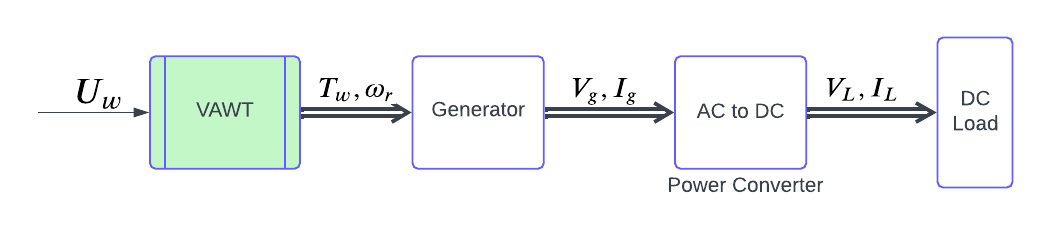}
	\caption{A schematic representation of the system}
	\label{fig:blockdia}
\end{figure}

The VAWT's aerodynamics, the power-electronic structure, the load and generator models are all discussed in detail in this part. This paper uses the model and settings presented in \cite{ugur} to simulate the VAWT system.
The block diagram representations, for the VAWT system and the control approach for WECS, are illustrated in Figures \ref{fig:WECS} \citep{augababaouglu2019bayesian} and \ref{fig:blockdia}.

Wind turbines use aerodynamics to transform the kinetic energy of the wind into mechanical energy, providing torque for a generator at a rotational speed $\omega_r$ and torque $T_{w}$. In this setup, a wind-powered generator produces electricity. The control mechanism typically measures $\omega_r$ and $T_{w}$ to derive reference load current as the required control effort. The amount of generated power depends on the wind's velocity $U_w$ and rotor's aerodynamics ($\rho~C_p S_a$ representing density of air, wind power coefficient, and wind turbine's swept area using rotor radius $r_{r}$ and length of blade $l_{b}$) according to equation \eqref{eq: power_wind} \citep{da2021fundamentals}.


\begin{equation}\label{eq: power_wind}
\mathcal{P}_{} = 0.5\rho~C_p S_a U^3_{w}\quad 
\text{with} \quad S_a = 2r_{r}l_{b}
\end{equation}
The conversion of the wind energy is limited by the aerodynamic efficiency of the rotor $C_p$. For a given wind speed, there is a rotor speed  $\omega_r$ such that the TSR is at the maximum of the $C_p$ curve. This operating point ensures maximum aerodynamic efficiency,
which is the main objective of most studies devoted to improving the energy efficiency of wind turbines. There is a minimum wind speed below which the rotor has too low efficiency and cannot start rotating, and a maximum wind speed where the centrifugal loads would cause damage and it is not allowed to reach by mechanical brakes.

\subsection{VAWT Parameters and Mathematical Model}

\begin{figure}
	\centering
	\includegraphics[scale=0.5]{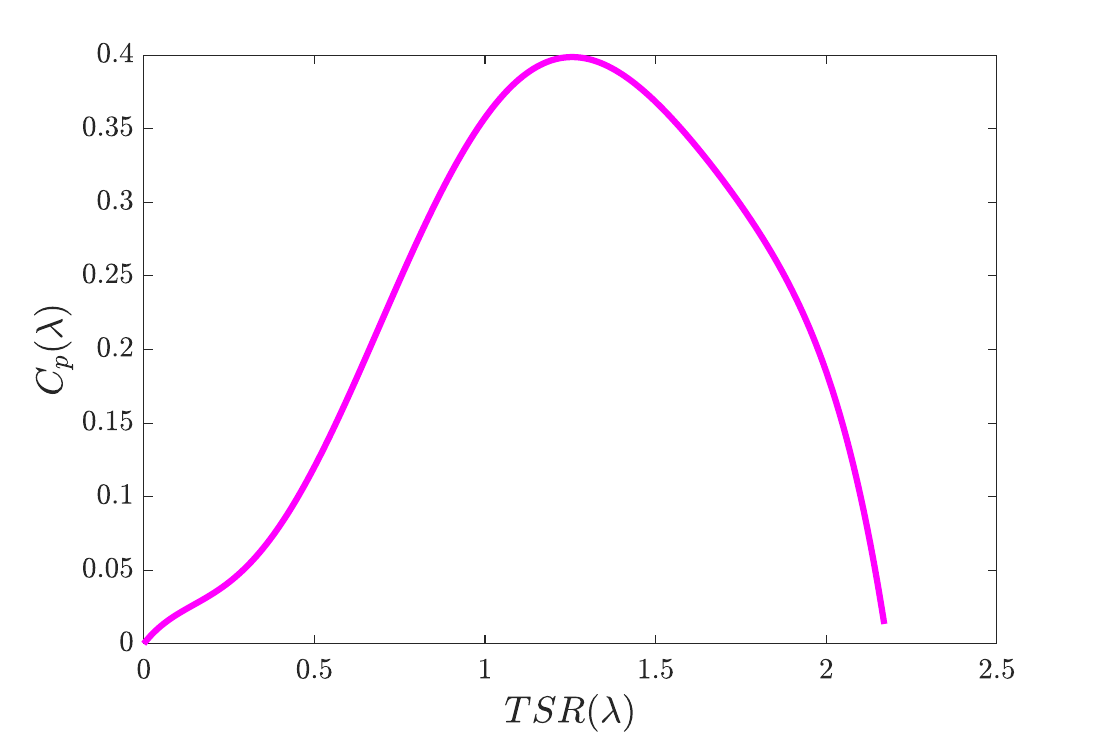}
	\caption{$\lambda - C_p$ curve of studied system}
	\label{fig: Cp curve}
\end{figure}

It should also be noted that $C_{p}$ itself is a function of TSR ($\lambda$) that is given in equation \eqref{eq: TSR}:

\begin{equation}\label{eq: TSR}
\lambda = \frac{\omega_{r}r_r}{U_{w}}
\end{equation}
The TSR, which provides a measurement of the ratio between the wind speed and the speed of the blade tip, is crucial in establishing the maximum output power of a wind turbine. A greater TSR means that the blade is moving faster than the wind, which typically results in more power being produced. The efficiency with which a wind turbine transforms wind energy into mechanical energy is measured by $C_p$ on the other hand. It is defined as the ratio of the mechanical power output of the turbine to the total wind power incident on the rotor: As shown in Figure \ref{fig: Cp curve}, for a given wind turbine, the $C_{p}$ curve shows that the maximum $C_{p}$ occurs at a TSR that is near the optimal TSR for maximum power output. The shape of the $C_{p}$ curve depends on the design of the wind turbine and can be used to compare different turbine designs and thus optimize the design of new turbines.
In order to simulate the behaviour of the wind and VAWT, equation \eqref{eq: P_wind}, which is the extension of equation \eqref{eq: TSR}, is used. These equations give the maximum possible value 
for $C_{p}$ as around $0.4$, as also seen from Figure \ref{fig: Cp curve}. For our setting, the VAWT parameter values are provided in Table \ref{tab:VAWT_sys}. Besides these structural parameters, a $6^{th}$order nonlinear relationship between $C_P$ and TSR is experimentally established as given in equation \eqref{eq: Cp_eq}.
\begin{equation}\label{eq: P_wind}
\mathcal{P} = \rho C_{p}\left(\lambda\right) r_{r} l_{b} U^3_{w}
\end{equation}

\begin{table}
	\begin{center}
		\caption{Parametrization of the system model.}
		\label{tab:VAWT_sys}
		\begin{tabular}{|l|c|cc|} 
			\hline
			\hline
			\quad \textbf{Description} & \textbf{Symbol} &  \textbf{Quantity} &\\
			\hline
			\hline
			\quad Rotor inertia & $J_r$ & $2$&$kg.m^2$\\
			\quad Rotor radius & $r_{r}$ & $0.5$&$m$\\
			\quad Blade length & $l_{b}$ & $1$&$m$\\
			\quad Friction coefficient & $b_{fr}$ & $0.02$&$Ns/rad$\\
			\quad Air density & $\rho$ & $1.2$&$kg/m^3$\\
			\hline
		\end{tabular}
	\end{center}
\end{table}
\begin{subequations}
	
	\begin{equation}\label{eq: Cp_eq}
	C_p\left(\lambda\right) = p_1\lambda^6 + p_2\lambda^5 + p_3\lambda^4 + p_4\lambda^3 + p_5\lambda^2 + p_6\lambda
	\end{equation}
	\begin{equation}\label{eq: Cp_param}
	p_{i} = [-0.3015,~1.9004,~-4.3520,~4.1121,~-1.2969,~0.2954]\quad \text{for}\quad i = 1, \dots, 6   
	\end{equation}
\end{subequations}
Equation \eqref{eq: Cp_param} provides different $C_{p}$ values and in Figure \ref{fig: Cp curve} a curve representing $C_{p}$ with respect to $\lambda$ has been shown.
One can also obtain generated torque by the wind by having the ratio between wind power and rotor velocity as in equation \eqref{eq: wind_torque}.  

\begin{equation}\label{eq: wind_torque}
T_{w} = \frac{P_{w}}{\omega_r}   = \frac{\rho C_p\left(\lambda\right)r_r l_b U_{w}^3}{\omega_r} 
\end{equation}

\subsection{Three-Phase PMSG-Rectifier's Model in the VAWT Structure}
\begin{figure}
	\centering
	\includegraphics[scale=0.4]{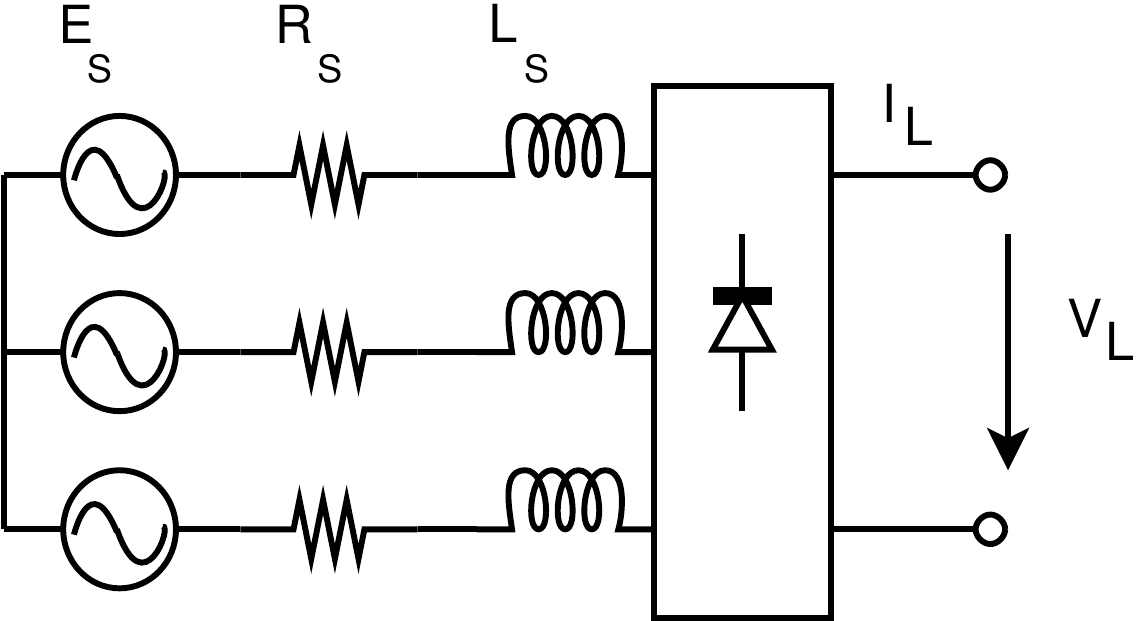}
	\caption{Three-phase PMSG-Rectifier schematic.}
	\label{fig: PMSG_rect}
\end{figure}  
Dynamical equation for the rotor's PMSG can be written as equation \eqref{eq: eq_montion} according to \cite{TRIPATHI2015PMSG} where $T_{g}$ and $T_{fr}$ are generator and friction torques (equations \eqref{eq: Tgen}-\eqref{eq: friction_torque}), respectively.
\begin{equation}\label{eq: eq_montion}
J_r \frac{d\omega_r}{dt} + T_{fr} + T_{g} - T_{w} = 0   
\end{equation}
\begin{equation}\label{eq: Tgen}
T_{g} = \mathcal{K}_t I_{L}
\end{equation}
\begin{equation}\label{eq: friction_torque}
T_{fr} = b_{fr} \omega_{r}  
\end{equation}

Figure \ref{fig: PMSG_rect} is an illustration of the PMSG-rectifier circuit. In this figure, $L_{S}$, $R_{S}$, and $E_{S}$ represent inductance, resistance and electromotive force for the PMSG, respectively. 

Load voltage $V_L$, can be calculated as shown by equation \eqref{eq: Vdc}, taking into account both the load current $I_L$ and $\omega_r$ in which a zero $I_L$ will bring the $V_L$ to its highest value. Because of the presence of $T_g$, the more one increases $I_L$, the more decrease will occur for $V_L$, as well.
\begin{equation}\label{eq: Vdc}
V_{L} = \sqrt{E_{SDC}^2 + \left(p\omega_rL_{dc}I_{L}\right)^2}-(R_{dc}+R_{D})I_{L}
\end{equation}

\begin{figure}
	\centering
	\includegraphics[scale=0.4]{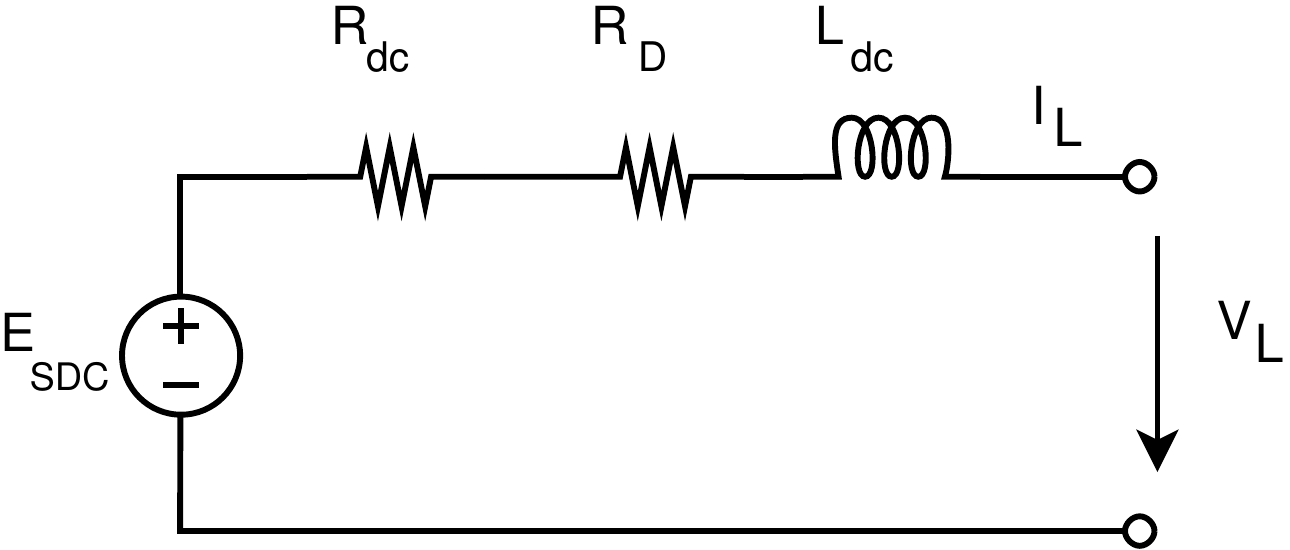}
	\caption{PMSG-Rectifier's modified DC model.}
	\label{fig: Simp_dc}
\end{figure}

Figures \ref{fig: PMSG_rect} and \ref{fig: Simp_dc} illustrate, respectively, the PMSG-rectifier model and its corresponding DC circuit, where their parameters can be specified according to Table \ref{tab:PMSG}. The PMSG-Rectifier voltage drops are due to both $R_{dc}$ and $R_D$ (equation \eqref{eq: R_over}) resistors, where the latter is used to design a more practical DC structure based upon the mean of the voltage drops associated with the generator's rotor reaction, commutating and overlapped currents in the three-phase diode bridge.
\begin{equation}\label{eq: R_over}
R_{D} = \frac{3L_sp\omega_r}{\pi}
\end{equation}

\begin{table}
	\begin{center}
		\caption{PMSG-Rectifier parameters.}
		\label{tab:PMSG}
		\begin{tabular}{|l |c |r|} 
			\hline
			\hline
			\textbf{Description} & \textbf{Equivalent PMSG} & \textbf{Simplified DC Model}\\
			\hline
			\hline
			Flux  & $\phi_s$ &  $\phi_{dc} = 3\sqrt{6}\phi_s/\pi$ \\
			EMF & $E_s = \phi_sp\omega_r$  & $E_{SDC} = 3\sqrt{6}E_s/\pi$\\ 
			Inductance & $L_s$ & $L_{dc} = 18L_s/\pi^2$ \\
			Resistance & $R_s$ & $R_{dc} = 18R_s/\pi^2$ \\
			\hline
			
		\end{tabular}
		
	\end{center}
\end{table}

\subsection{The Load Model of VAWT}
\begin{figure}
	\begin{center}
		\includegraphics[scale=0.4]{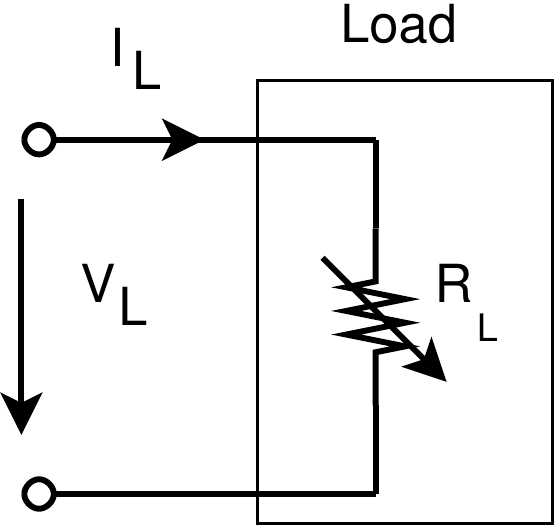}
		\caption{Load controlled by optimization algorithm\label{fig: Simp_load}}
	\end{center}
\end{figure}

In this study a simplified load circuit consisting of a variable load 
resistor
is used as depicted in Figure \ref{fig: Simp_load}. Adjusting $R_L$ to increase $I_L$, will produce a large load-torque to the turbine mechanics; alternatively, decreasing $I_L$, will likewise decrease the generator torque. The RL-MCMC algorithm is thus responsible for regulating $I_L$ so that the $\omega_r$ may be adjusted in order to achieve the highest possible output energy. 

\section{Structure of the RL and MCMC}\label{sec: RL and MCMC}
As the controller (agent) in the RL setting iteratively via a decision-making process interacts with the environment in which they operate, the system may learn the optimal policy behavior. It is through the selection of actions that an agent may gain knowledge from the experiences it has had. The agent gets an observation for per step $t$, executes an action and moves to a new state in its environment, and is rewarded with a real-valued number ${r}_{t} \in \mathbb{R}$; in the long run, RL seeks to discover a policy that maximizes this value. The policy may be thought of as a function that connects observations with the appropriate actions. The main components of an RL problem can be formulized using the following set:
\[
({S}, {A}, \xi, \vartheta, {r})
\]
For representing set of continuous state and action spaces separately, we use ${S}$ and ${A}$ where each individual state-action for a given time step $t$ belongs to these spaces.
For each $t$, the agent's future state ${s}_{t+1}$ is determined by a transition dynamic distribution $\xi(.)$ having the present state-action of the agent (${s}_{t}$, ${a}_{t}$), where $\vartheta(.)$ is the initial state function.

\[
\xi({s}_{t+1} | {s}_{t}, {a}_{t}) \quad \forall {s}, ~{s}_{t+1}\in {S}, \quad \forall {a}_{t} \in {A}
\]
The policy is indicated by a parameterized ($\theta \in \mathbb{R}^{d_{\theta}}$) distribution ${h}_{{\theta}} ({a}_{t} | {s}_{t})$ where they can be chosen from a stochastic \emph{Gaussian} density, allowing for a proactive exploration of the state-space. Assuming ${X}_{t} = ({S}_{t}, {A}_{t})$, forms a Markov chain state-action trajectory for $\{ {X}_{t} \}_{t \geq 1}$ with a transition law given in \eqref{eq: Markov chain transition density} where state-action pair for time $t$ is defined as $x_{t} = (s_t, a_t)$.
\begin{equation} \label{eq: Markov chain transition density}
{f}_{\theta}(x_{t+1} | x_{t}) := \xi({s}_{t+1} | {s}_{t}, {a}_{t}) {h}_{\theta}({a}_{t} | {s}_{t}).
\end{equation}
To have a measure of the the conformity of the state transitions for a whole trajectory, weighted rewards' summation ($\gamma \in (0,1]$) over a time $T$ is calculated; namely called \emph{return}
\begin{equation}\label{eq: return}
{R}_{}(x_{1:T}) = {\sum}_{t = 1}^{T}\gamma^{t-1}r_{t}.
\end{equation}
Moreover, the joint distribution of the trajectory  $x_{1:T}$ can be given as, 
\begin{equation} \label{eq: trajectory distribution initial}
{p}_{\theta}(x_{1:T}) = \vartheta({s}_{1}) {\prod}_{t = 1}^{T} {f}_{\theta}(x_{t+1} | x_{t})
\end{equation}
with $\vartheta({s}_{1})$ implying an initial state distribution in which ${s}_{1} \sim \vartheta(.)$. Having the trajectories' distribution and return, a cost function ${J}(\theta)$, evaluating the police's performance, can be defined, 
\begin{equation} \label{eq: J theta}
{J}_{}(\theta) = \mathbb{E}_{\theta}\big[{R}_{}(x_{1:T})\big] = \int {p}_{\theta}(x_{1:T}) {R}_{}(x_{1:T}){d} x_{1:T}.
\end{equation}

In RL, optimizing policy is its main purpose to finding ideal parameters $\theta^{\star}$. This could be accomplished by maximizing the return expectations over a trajectory: 
\begin{equation}
\theta^{\star} = \arg \max_{\theta \in \Theta} {J}(\theta)
\end{equation}

It is important to recognize that the trajectory's density in \eqref{eq: trajectory distribution initial} might be complicated or even partially known. Thus for such a density function, the integral must be addressed in equation \eqref{eq: J theta}, which is numerically insoluble. \cite{Levine:2016,Deisenroth2011,PETERS2008} have come up with several cutting-edge gradient-based RL algorithms in order to cope with this issue. However, one major challenge with the gradient-based methods is the local optimal solution. Opposed to these methods, Bayesian approaches are used to explore the possible high reward areas of the parameter space $\Theta$ in seeking optimal policy parameters as shown in \cite{PautratCM18,marco_ICRA_2017BO,Gal:2016}. Inspired by the $\emph{exploration-exploitation}$ dilemma in RL, problems related to the gradient-based RL, and based on the existing gap in applying Bayesian optimization methods in WECS, we provide an RL-based Bayesian method with an RBFNN controller which is capable of being applied to energy systems.

\subsection{RL-MCMC applied to the Dynamical Systems}
\subsubsection{An Overview to RL-MCMC}
Proposed RL-MCMC algorithm, combines the main ideas of the RL to learn optimal policies through decision making considering maximization of a long-term reward function. On the other side, Bayesian MCMC sampling often concentrates on sampling from complex distributions whenever dealing with numerical methods is unfeasible.  For the cases where the environment is either complex or involves uncertainty, the MCMC method can play a crucial role by generating sample trajectories from a posterior function of the RL algorithm. This posterior function can capture the uncertainties in the optimal policy given the state-action and reward trajectories. The posterior distribution can represent the RL algorithm's return function and be approximated using MCMC sampler. The Bayesian MCMC then guides the searching parameter space towards the high-reward regions of the posterior.
\subsubsection{RL-MCMC Application}
The suggested method, which is applicable to systems with continuous domains, is an MCMC-based policy search algorithm based on RL to control the WECS. In comparison to gradient-based RL, the Bayesian-MCMC offers a number of significant benefits. Its mathematical simplicity, along with the fact that it is not reliant on gradient computations, enables it to avoid being mired in locally optimal solutions.
Since the calculation of ${J}(\theta)$, entails prohibitive computational effort, we propose forming a density function that includes the policy parameters' prior distributions $\mu(\theta)$. Then, assuming the Markovian ergodicity property of parameters, the constructed density function, aka \emph{posterior}, is sampled by the MCMC algorithm, where expectation calculations are hard to achieve 
\begin{equation}\label{eq: poli_dist}
{\pi}(\theta) \propto \mu(\theta) {J}(\theta)
\end{equation}

We would like to clarify that the prior knowledge that is in question regards the policy parameter that determines the action given a state, and not the discounted reward. By “model free”, we actually meant that the dynamics of the system may not be analytically available. However, the action does depend on the state of the system through a vector of parameters, that is, the policy parameters denoted by $\theta$. As shown by \cite{Ghavamizadeh2016}, one of the main advantages of Bayesian RL is that it can benefit from prior information on the problem to help direct the sampling, for our case towards the areas with high rewards, to cope with the exploration- exploitation of the search space. By casting the RL problem as a Bayesian inference problem with the posterior distribution $\pi(\theta)$. We are thus able to embed any prior knowledge on what the policy parameter should be in the prior $\mu(\theta)$.

In MCMC, the target distribution is created accepting two properties of the chain as \emph{invariance} and \emph{ergodicity}, which is being initialized with a given  $\theta^{(0)}$. As an MCMC algorithm, Metropolis-Hastings (MH) selects the appropriate candidate parameter from a distribution of proposals i.e.\ $\theta^{\dagger} \sim \Gamma(\theta^{\dagger} | \theta)$. The suggested sample is then either taken with an acceptability degree given in \eqref{eq: accept-reject}, in which case the value of the current parameter is substituted with the new one, or dismissed, in which case the current parameter remains unchanged. 

\begin{equation}\label{eq: accept-reject}
\varrho(\theta, \theta^{\dagger}) = \frac{\Gamma(\theta | \theta^{\dagger})}{\Gamma(\theta^{\dagger} | \theta)} \frac{\mu(\theta^{\dagger}) {J}(\theta^{\dagger})}{\mu(\theta) {J}(\theta)}
\end{equation}
An analytical calculation appears to be impossible due to the presence of ${J}(\theta)$ and ${J}(\theta^{\dagger})$ in \eqref{eq: accept-reject}. 
However, an estimator with the property of obtaining unbiased and non-negative approximations of $J(\theta)$, enables us to sample from $\pi(\theta)$. This is generally possible by using an importance sampling (IS) problem. One of the improvements of the proposed RL-MCMC algorithm is mainly based on the structure of the total reward function, where instead of using an additive one, we propose to use a multiplicative total reward based on the idea of risk-sensitivity. According to \cite{Aviv2015}, the risk-sensitivity can handle the uncertainty involved in the environment. In this regard, Equation \ref{eq: J theta} can then be modified and written as:
\begin{equation}
{J}_{}(\theta) = \mathbb{E}_{\theta}\big[{R}_{}(x_{1:T})\big] = \int {p}_{\theta}(x_{1:T}) \prod_{t=1}^{T} \exp\left( \gamma^{t-1}r_{t}\right) {d} x_{1:T}.
\end{equation}
Summarizing the learning steps, we can give the pseudocode of the proposed RL-MCMC algorithm for WECS in Algorithm \ref{alg: Pseudo-marginal Metropolis-Hastings for reinforcement learning}. 
\begin{algorithm}
	\caption{RL-MCMC learning mechanism for WECS}
	\KwIn{Initialize policy parameters and estimated cost $\big(\theta^{(0)} = [\sigma^{(0)} \quad w^{(0)}], {J}^{(0)}\big)$}
	\KwOut{Policy parameters $\theta^{(\ell)}$, $\ell = 1, 2, \ldots$}
	\For{$\ell$ = 1, 2, $\ldots$}
	{sample a candidate parameter $\theta^{\dagger} \sim \Gamma(\theta^{\dagger} | \theta)$. \\
		Simulate WECS by proposed parameter $\theta^{\dagger}$ and calculate the state-action ${x}_t = ({s}_t, {a}_t) = (\dot{e}, I_{L_{ref}})$\\
		Calculate the reward ${r}_t = - {{s}_t}^2Q$\\
		Cost estimation using standard IS: ${J}(\theta^{\dagger})= {e}^{\big({\sum}_{t=1}^{T}{r}_t  \big)}$\\
		Choose candidate by a probability of $\min \big\{ {\varrho}(\theta, \theta^{\dagger}), 1 \big\}$, put $\theta^{(\ell)} = \theta^{\dagger}$ and ${J}^{(\ell)} = {J}^{\dagger}$ 
		\[
		{\varrho}(\theta, \theta^{\dagger}) = \frac{\Gamma(\theta | \theta^{\dagger})}{\Gamma(\theta^{\dagger} | \theta)} \frac{\mu(\theta^{\dagger})}{\mu(\theta)} \frac{{J}^{\dagger}}{{J}},
		\]
		else candidate is rejected, $\theta^{(\ell)} = \theta$ and ${J}^{(\ell)} = {J}$.
	}
	\label{alg: Pseudo-marginal Metropolis-Hastings for reinforcement learning}
\end{algorithm}

Based on the explanations brought up and the given algorithm and for the sake of clarification of the methodology, a step by step procedure is outlined as below:
	\begin{enumerate}
		\item Specify the state and action spaces, the reward function and the initial policy parameters for the RL problem.
		\item Construct the posterior distribution by defining prior densities and the cost function of the RL.
		\item Define a suitable MCMC sampler to generate samples which for our case we use a Metropolis-Hastings sampler.
		\item Assign a proposal distribution for the MCMC and draw candidate parameters.
		\item Collect sequence of state-action-reward tuples from the WAVT system.
		\item Estimate the cost function and then based on the acceptance probability either accept or reject the sample proposal.
		\item Update the policy parameters using a Gaussian random walk.
		\item repeat $4-7$ until a sufficient number of samples have been generated to estimate the posterior distribution.
	\end{enumerate}
	
\section{Proposed Control Structure}\label{sec: proposed_control}
Our goal is to create a framework that along with the ability to control VAWT's system with unknown dynamics can also obtain desirable responses towards different wind velocity profiles. In the present part, we will cover the RL environment, the RBFNN controller architecture, and different steps to train the proposed RL- MCMC algorithm for the WAVT. To improve the overall electrical energy output of the WAVT system, the instantaneous current of the load $I_L$ in the generator will be optimized provided that the current and voltage in the load do not exceed their their maximum values. For this purpose, an RBFNN controller is used to compute the corresponding current in the load as a reference $I_{L_{ref}}$, where the proposed control structure can be feasible for the RL-MCMC algorithm in learning the unknown parameters of the VAWT.

%
%
The instantaneous electrical power can be integrated within a given interval, to determine the desired energy output using the formula \eqref{eq: energy_gen}. The ideal aerodynamic power is produced by the rotor upon continually keeping $C_{p}$ at its highest amount, and integrating over it yields the maximum reference mechanical energy, which is then converted to an optimal electrical energy value \citep{ugur} which results in \eqref{eq: energy_ref}. After calculating ${E}_{out}$ and ${E}^{\star}$, the resulting energy error, which its derivative will be used as the required state of the learning algorithm, can simply be obtained as \eqref{eq: error}.

\begin{equation}\label{eq: energy_gen}
{E}_{out} = \int_{0}^{\tau}{P} dt
\end{equation}
\begin{equation}\label{eq: energy_ref}
{E}^{\star} = \int_{0}^{\tau}{P}^{\star}dt
\end{equation}

\begin{equation}\label{eq: error}
e = {E}^{\star} - {E}
\end{equation}
Similar to the state of the learning algorithm $\dot{e}$, a continuous action $I_{L_{ref}}$ with a single dimension is considered.

\subsection{RBFNN as the Controller}\label{sec:Structure of RBNN} 
A nonlinear RBFNN control given in \eqref{eq: RBF_out} is designed to calculate the reference load current $I_{L_{ref}}$ with $n$ hidden nodes.

\begin{equation}\label{eq: RBF_out}
F\left(x,\theta\right) = \sum_{i = 1}^{n}w_i\mathcal{R}_i\left(x\right) + \mathrm{b}
\end{equation}
where, $\theta$ shows the adjustable parameters of the system: $\theta=[w \, \,  \sigma]$.

The Gaussian receptive field $\mathcal{R}_i\left(x\right)$ with input $x$ as shown in Table \ref{tab:RBNN_parm} is defined in \eqref{eq: RBF} (with centers $c_{ij}$ and variance $\sigma^{2}_{j}$), calculates $i^{th}$ node output, $b$ is a biasing scalar value, and $w_{i}$s represent the weights.
%
%
\begin{equation}\label{eq: RBF}
\mathcal{R}_i\left(x\right)=\sum_{j=1}^{m} \exp\left(-\frac{\|{\left(x_j-c_{ij}\right)\|}^2}{2\sigma_j^2}\right) 
\end{equation}

\begin{subequations}\label{eq: RBF center}
	\begin{equation}
	c_{ij} = \begin{bmatrix} 
	c_{11} & \hdots & c_{1m} \\
	\vdots & \ddots & \vdots\\
	c_{n1} &  \hdots & c_{nm} 
	\end{bmatrix}
	\end{equation}
	\begin{equation}\label{eq: RBF width}
	\sigma = \begin{bmatrix} 
	\sigma_1 &\sigma_2& \hdots &\sigma_m
	\end{bmatrix}^T
	\end{equation}
\end{subequations}

In our model, $c_{ij}$ matrix is defined by taking the relevant interval of related variable and dividing them into $5$ equally spaced intervals, each of which contains an RBF function. Table \ref{tab:RBNN_parm_val} lists the boundary points of the intervals where they are selected according to the working region of the inputs of RBFNN. 
The inputs for the RBFNN are given as in Table \ref{tab:RBNN_parm} which is arranged taking into account the probable wind profiles as well as the internal dynamics of the VAWT. In this regard, the physical parameter $U_w$ and its derivative ${\dot{U}_{w}}$, are chosen as the primary means through which wind velocity and its rate of change may be perceived by RBFNN. On the other hand, VAWT's internal states such as a $V_L$, $I_L$, $\omega_r$, and $\dot{\omega}_{r}$ are fed into the neural controller.
\begin{table}
	\begin{center}
		\caption{RBFNN structure and its parameters}
		\label{tab:RBNN_parm}
		\begin{tabular}{|c|c|c|} 
			\hline
			\hline
			\textbf{Inputs for the RBFNN} & \textbf{Structural parameters}& \textbf{Parameter description}\\
			\hline
			\hline
			$x_1$  & $U_w$& Wind Speed\\
			$x_2$  & ${\dot{U}_{w}}$& Wind speed derivative\\
			$x_3$  &  $I_L$& Current for the load \\
			$x_4$  &  $V_L$& Voltage for the load\\
			$x_5$  &  $\omega_r$& Rotational velocity for PMSG\\
			$x_6$  &   $\dot{\omega}_{r}$& PMSG rotational velocity's derivative\\
			\hline
		\end{tabular}
	\end{center}
\end{table}

\begin{figure}
	\centering
	\includegraphics[scale=0.8]{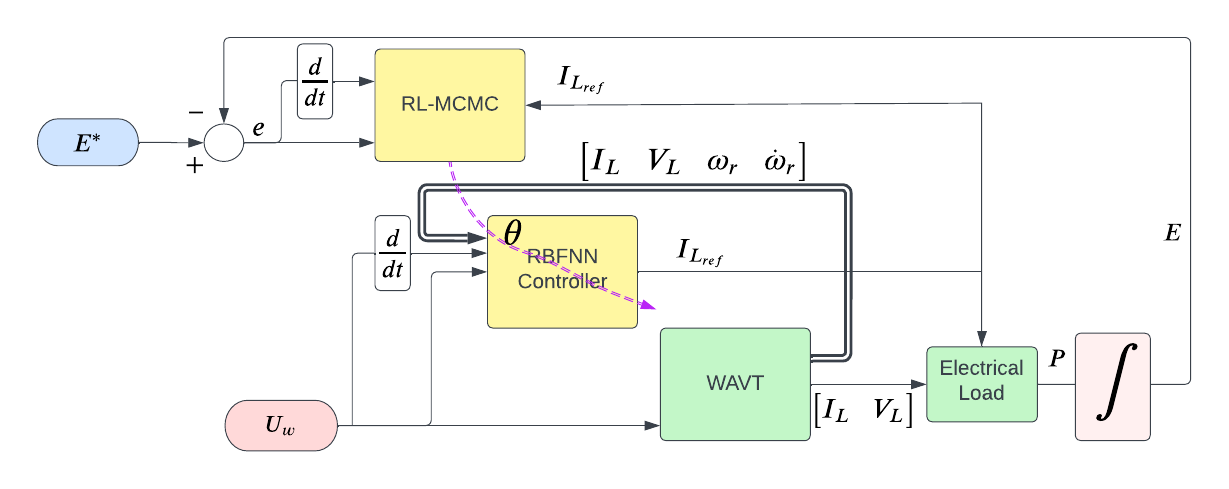}
	\caption{RL$-$MCMC learning block diagram for WAVT with the proposed RBFNN controller.}
	\label{fig:RBFNN}
\end{figure}

The proposed RL-MCMC approach is implemented to learn $\theta$ parameters of the controller. Taking advantage of such compound learning mechanism, an improved performance, when experiencing real winds, will be achieved via determining the ideal $I_{L_{ref}}$ for all possible windy situations. Moreover, it can help to learn an unknown model of the system by characterizing and encoding the model parameters into RBFNN structure. Refer to the diagram in Figure \ref{fig:RBFNN} to see how the suggested control mechanism is performing.  

\subsection{Training Procedure of the Learning Algorithm}\label{subsec: steps}
The section explains the way that the control parameters could be taught under intricate wind samples.
The training is begun by applying three different wind profiles as step, sinusoidal and realistic wind data with an initial parameter set $\theta_{S0}$ to learn the optimal parameter sets $\theta_{S1}$, $\theta_{S2}$. The training scheme is carried out as follow:

\begin{itemize}
	\item {\bf Step 1:} The parameters for the policy are initialized as $\theta_{S0}$ 
	\item {\bf Step 2:} In the first stage of training, a step wind signal is applied to the VAWT and  $\theta_{S1}$ is learned.
	\item {\bf Step 3:} During training's $2^{nd}$ stage, , a sine wind signal is applied to the VAWT and $\theta_{S2}$ is obtained.
\end{itemize}

The reason for selecting a step wind is to consider it as a simple signal which enables the rotor for an energy management strategy. Afterwards, training's $2^{nd}$ stage targets to learn a controller to be responsive to a wind profile with changing speed. Therefore, a sinusoidal wind with a close frequency to that of the realistic ones, is selected. In an actual use case, the system would be presented to the customer having completed up to Step 3 training, where it is partially but not completely optimized. After commissioning, the system adapts to the local wind patterns. 
Without initial training up to Step $3$, convergence to the best pattern can be impossible for the system within a reasonable time. 

\section{Simulated Experiments and Analysis}
\label{sec: results} 
To demonstrate the efficiency of the proposed RL-MCMC algorithm over the classical MPPT method, an extensive simulation procedure is done for two different MPPT structures as shown in Table \ref{tab:MPPT_desc}. In the first part of this section, numerical values for the parameters of the RL-MCMC structure and the RBFNN controller used in this section are given. 
In the second part, the results of each step of training as detailed in Sec. \ref{subsec: steps} of the proposed method are given.
After parameter initialization in Step 1, a step wind is implemented into the system in Step 2 and the learned policy parameters during training stages of the RL-MCMC and the resulting time response specifications of the VAWT ($P, \omega_{r}, V_{L}, I_{L}$, and $R_{L}$) are illustrated. In 
Step 3, a sinusoidal wind is applied and their corresponding time responses and learned parameters are provided. Finally,
in Step 4,
for a better comparison, a realistic wind profile with noise is applied to the system.

\subsection{Numerical values of RL-MCMC for Simulations}\label{subsec: 1}
For the proposed RL-MCMC algorithm the reward function is taken as,  
\begin{equation} 
{r}_t = -{s}_t^2 Q
\end{equation} 
with a state weight $Q = 10^5$ and state ${s}_t = \dot{e}$. We assume an average return function with $\gamma = 1$ in \eqref{eq: return}.

Policy parameters in the RL-MCMC, are updated according to a Gaussian random walk with a proposal density as $\Gamma(\theta^{\dagger} | \theta) = \mathcal{N}(\theta^{\dagger} ; \theta, \Sigma_{\Gamma})$ in which a covariance matrix $\Sigma_{\Gamma}$ is defined as $\mathrm{diag}\left(\left[\begin{matrix}
1 & \hdots&1
\end{matrix}\right]_{1 \times dim(\theta)}\right)$. The prior distribution $\mu(\theta)$, for policy is $\mathcal{N} \left (0; \mathrm{diag} \left (\left[\begin{matrix}
10^{4}& \hdots& 10^{4}
\end{matrix}\right]_{1 \times n_{\theta}}^{T} \right ) ,\Sigma_{\Gamma}\right)$, where $dim({\theta}) = 12$ is the total number of parameters to be learned. The sampling time of VAWT dynamics is $1~ms$. 
Also, RBFNN structure is shown in Table \ref{tab:RBNN_parm_val} with a bias value as $3.5$. It worth noting that if we have less definitive knowledge on where $\theta$ parameters might be, this can be incorporated into the posterior by assigning a smoother shape to prior $\mu(\theta)$ such as a Gaussian kernel. This is what we did in our experiments. However to reflect an uncertain parameter space for our case, an “uninformative” prior distribution is used.
\begin{table}
	\begin{center}
		\caption{Ranges  of RBFNN Inputs.}
		\label{tab:RBNN_parm_val}
		\begin{tabular}{|c|c|c|c|} 
			\hline
			\hline
			\textbf{RBFNN} & \textbf{Symbol} & \textbf{Min} & \textbf{Max}\\
			\textbf{Input}& & \textbf{Center} & \textbf{Center}\\
			\hline
			\hline
			$x_1$  & $U_w$ & 4.66 & 11.31\\
			$x_2$  & ${\dot{U}_{w}}$ & -8.33 & 8.32\\
			$x_3$  &  $I_L$ & 0.83 & 9.13 \\
			$x_4$  &  $V_L$ & 3.32 & 36.52 \\
			$x_5$  &  $\omega_r$ & 5& 35 \\
			$x_6$  &   $\dot{\omega}_{r}$ &-4.998 & 4.992 \\
			\hline
		\end{tabular}
	\end{center}
\end{table}
\subsection{RL$-$MCMC's $1^{st}$ Training Stage}\label{subsec: train1}
Since the examined VAWT's working region for $U_{w}$ belongs to $[6,~12]$~$\frac{m}{s}$, a wind as a step function of $8~\frac{m}{s}$ amplitude is applied. To obtain VAWT's dynamic performance, simulation time is selected as $150~s$. The vector of initial parameters $\theta_{S0}$ is defined as:
\begin{equation}
\theta_{S0} = \left[\begin{matrix} \sigma_{S0}&w_{S0}\end{matrix}\right]\nonumber
\end{equation}
where for $\sigma_{S0}$ and $w_{S0}$ we have $\frac{n_{\theta}}{2}$ parameters in which for each of them $\sigma_{S0} = 20$ and $w_{S0} = 1$ 



\begin{figure}
	\centering
	\begin{subfigure}[b]{0.48\linewidth}
		\includegraphics[width=\linewidth]{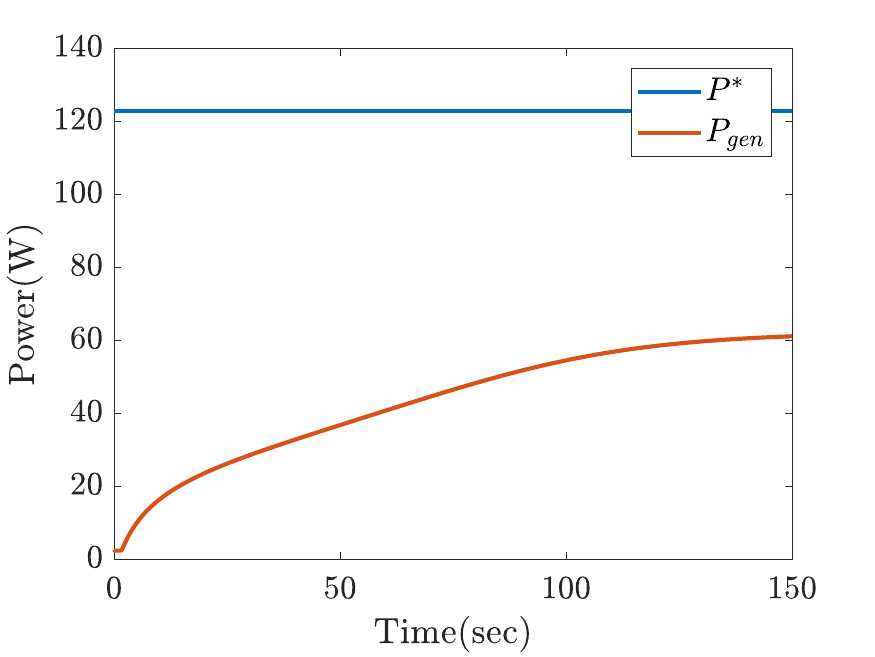}
		\caption{Nominal versus obtained mechanical power.}
			\label{subfig: T0_P} 
	\end{subfigure}
	\begin{subfigure}[b]{0.48\linewidth}
		\includegraphics[width=\linewidth]{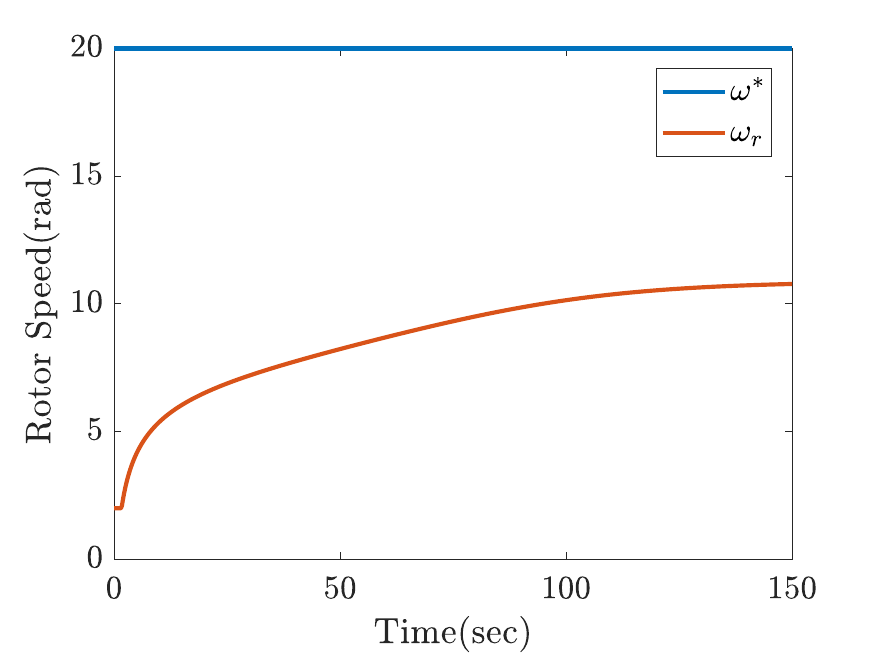}
		\caption{Rotor's angular velocity.}
		\label{subfig: T0_W}
	\end{subfigure}
	\begin{subfigure}[b]{0.48\linewidth}
		\includegraphics[width=\linewidth]{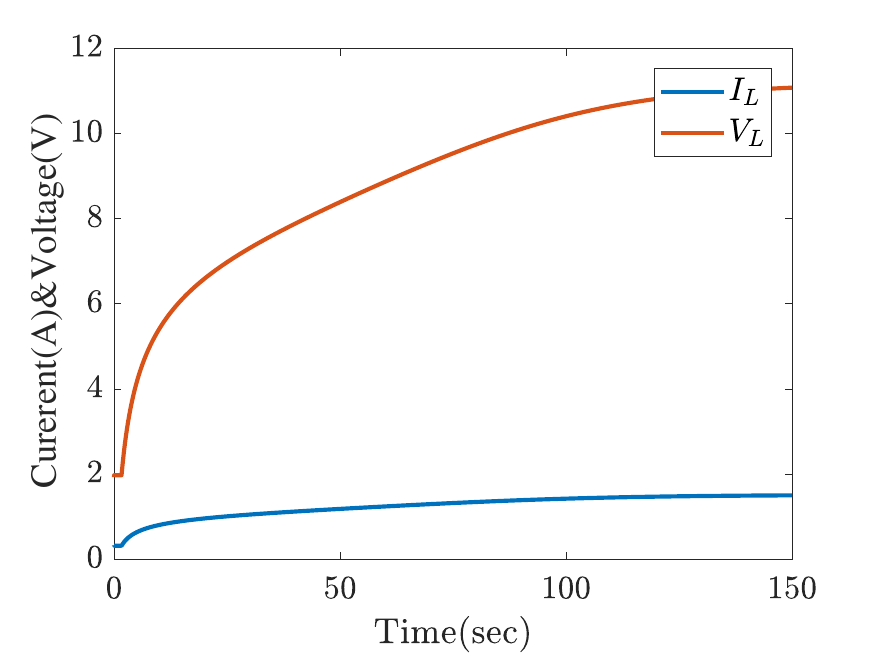}
		\caption{Measured current and voltage in the load.}
	\end{subfigure}
	\begin{subfigure}[b]{0.48\linewidth}
		\includegraphics[width=\linewidth]{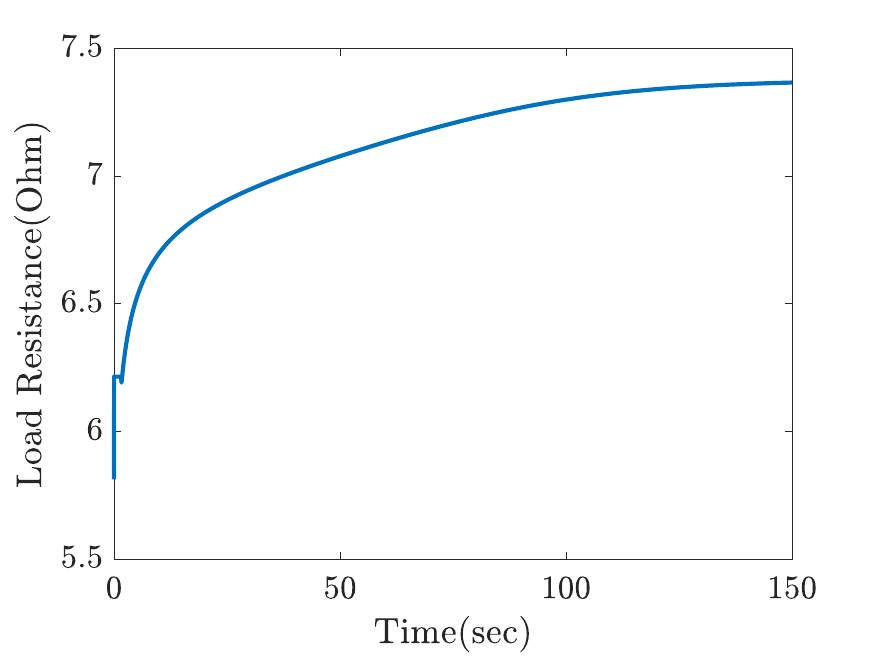}
		\caption{Resistance of the load.}
	\end{subfigure}
	\caption{Time response results for $P$, $\omega_{r}$, $V_L$, $I_L$, and $R_L$ by RL-MCMC 
		using initial parameters $\theta_{S0}$.}
	\label{fig: 1_stage_initial_sim}
\end{figure}
The response of the system before learning takes place
using $\theta_{S0}$, is illustrated in Figure \ref{fig: 1_stage_initial_sim}. As expected and can be seen from Figures \ref{subfig: T0_P}, and \ref{subfig: T0_W}, the system exhibits inferior performance. These figures represent the mechanical power compared to nominal power and rotor angular velocity compared to its nominal value, respectively. 
Figure \ref{fig: 1_stage_initial_sim} is displayed here only as a baseline. The poor performance is mainly due to the fact that the algorithm is in its initial state and struggles to learn more about the search space of the parameters. Therefore, the more training stages, the better becomes the learning performance of the RL-MCMC  algorithm.

\begin{figure}
	\centering
	\begin{subfigure}[b]{0.49\linewidth}
		\includegraphics[width=\linewidth]{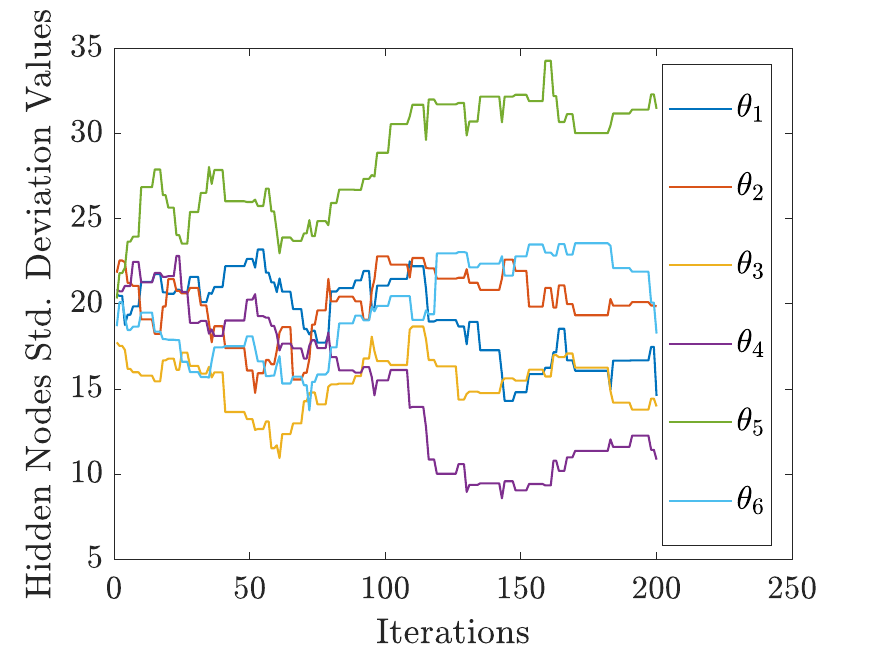}
		\caption{~Learned standard deviation.}
		\label{subfig: std}
	\end{subfigure}
	\begin{subfigure}[b]{0.49\linewidth}
		\includegraphics[width=\linewidth]{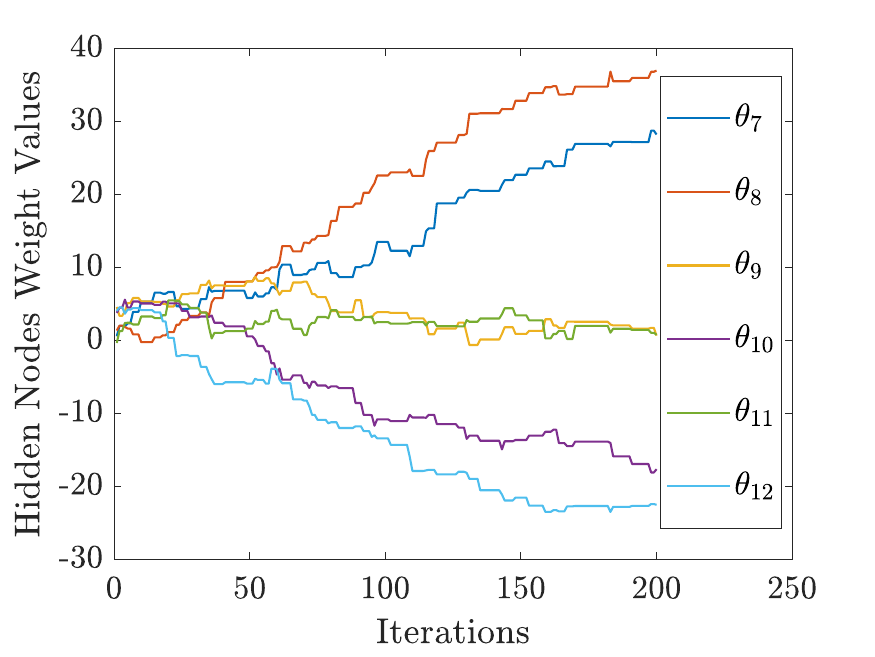}
		\caption{~Learned weights.}
		\label{subfig: weight}
	\end{subfigure}
	\caption{Trace plots of the parameters during the first stage of training.}
	\label{fig: train1_output}
\end{figure}

The evolution of policy parameters 
during the first stage of training are shown in Figure \ref{subfig: std}, and \ref{subfig: weight}. 
The first-stage training parameters ($\theta_{S1}$) are generated by taking the average of sample parameters after the parameters have achieved a stable distribution. In our case the average value of each parameter in the last 50 iterations is taken as its final value.

\begin{figure}[t!]
	\centering
	\begin{subfigure}[b]{0.49\linewidth}
		\includegraphics[width=\linewidth]{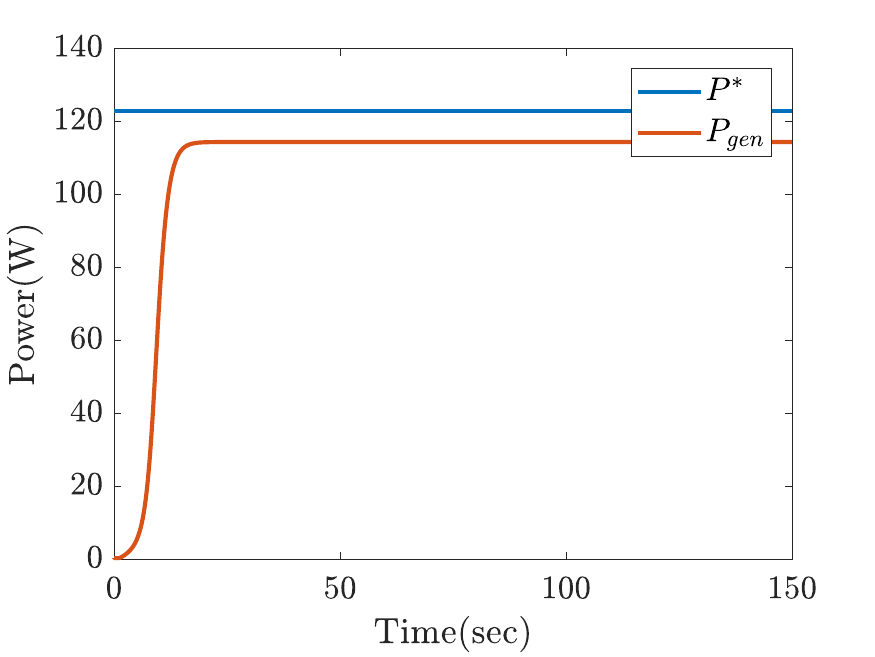}
		\caption{Nominal versus obtained mechanical power}
		\label{subfig: mech_pow_step}
	\end{subfigure}
	\begin{subfigure}[b]{0.49\linewidth}
		\includegraphics[width=\linewidth]{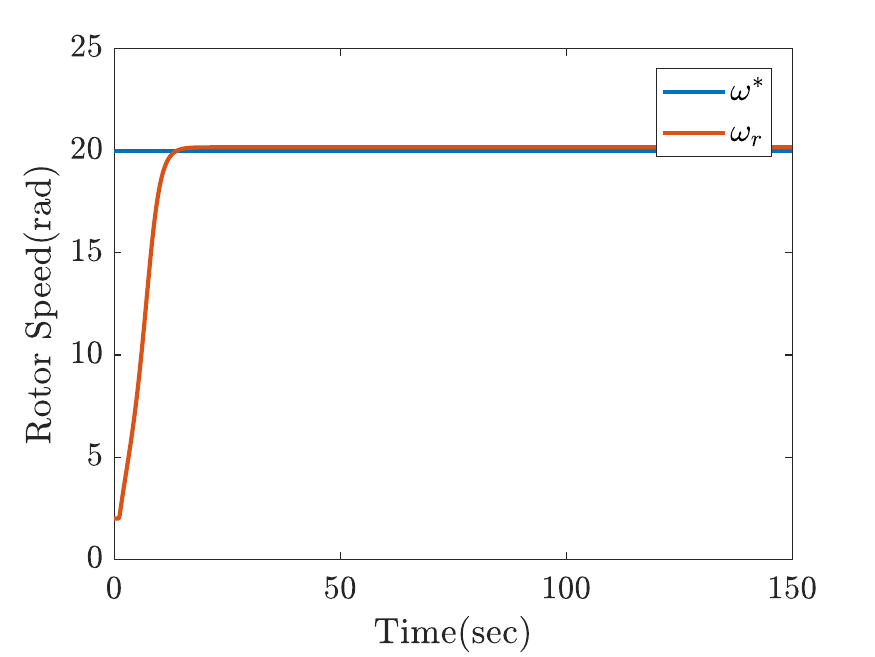}
		\caption{Rotor's angular velocity}
		\label{subfig: gen_rotor_speed}
	\end{subfigure}
	\begin{subfigure}[b]{0.49\linewidth}
		\includegraphics[width=\linewidth]{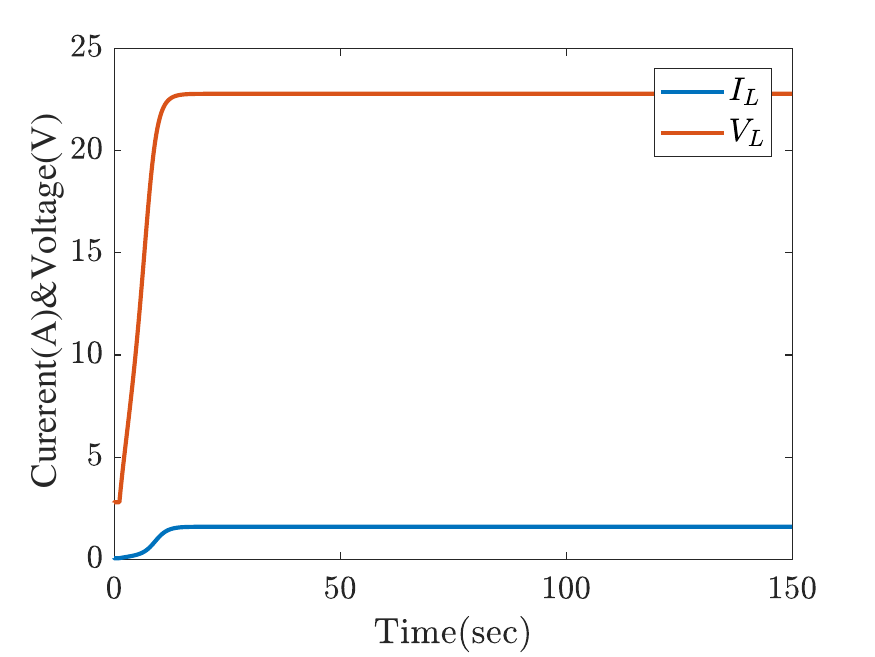}
		\caption{Measured current and voltage in the load}
		\label{subfig: load_curr_volt}
	\end{subfigure}
	\begin{subfigure}[b]{0.49\linewidth}
		\includegraphics[width=\linewidth]{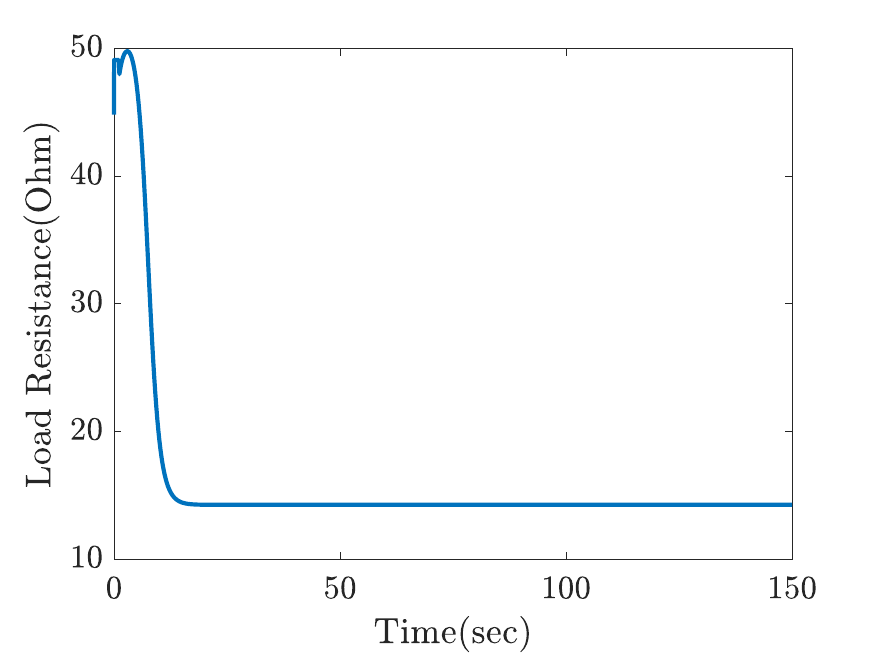}
		\caption{Resistance of the load}
		\label{subfig: load_rezz}
	\end{subfigure}
	\caption{Time response results for $P$, $\omega_{r}$, $V_L$, $I_L$ and $R_L$ by RL-MCMC  after first stage training, using $\theta_{S1}$ parameters.}
	\label{fig: 1_stage_final_sim}
\end{figure}

The time response is a more intuitive result, and is shown in Figure \ref{fig: 1_stage_final_sim}. The performance of the system has improved compared to the parameter set $\theta_{S0}$ and it can effectively deal with the applied wind. However, the performance is still far from optimal due to lack of richness in the training set. In Figure \ref{subfig: mech_pow_step}, the optimal mechanical power (from the wind) is compared to mechanical power obtained from the generator. Mechanical rather than electrical power is considered to avoid inserting the generator efficiency into the discussion. In Figure \ref{subfig: mech_pow_step} it is observed that the power did not reach the optimum value. Longer training might increase the model's precision and possibly bring the power to its optimum level. The model's performance, however, could also be influenced by additional variables, such as the model's architecture, the hyperparameters chosen, and the quality of the training data.

The resultant RL-MCMC controller produces a rotor speed for the generator that is close to the optimal value, as shown in Figure \ref{subfig: gen_rotor_speed}. More importantly, the proposed control algorithm does not exhibit the type of harsh rise for the current as seen in Figure \ref{subfig: load_curr_volt}, but instead, there is a delay before the current increases. This delay is important because it allows a light load on the rotor, allowing it to accelerate to optimum speed $\omega_r$ quickly. In e.g. an MPPT controller, a step wind would produce an immediate rise in the current which is a greedy approach that maximizes instantaneous power but delays the acceleration of the rotor to the optimal speed due to increased mechanical load, thus reducing the total energy output of the system.

The load resistance graph given in Figure \ref{subfig: load_rezz}, shows the ratio $V_L / I_L$, (depicted in Figure \ref{subfig: load_curr_volt}). It is clearer to see from Figure \ref{subfig: load_rezz} that at the beginning the proposed control method decided to draw a small amount of power during the spooling time of the rotor so that it reaches a value close to optimum. After that, it lowers the load resistance so that more power is drawn. This trait shows that the algorithm is already exploiting, at this stage, the fact that it is more advantageous to allow the rotor to get up to optimal speed quickly before applying load. Greedy algorithms such as MPPT try to extract power as soon as rotor speed starts to increase and therefore have worse energy output.
\subsection{RL-MCMC's $2^{nd}$ Training Stage}\label{subsec: train 2}
\begin{figure}
	\centering
	\includegraphics[scale=0.5]{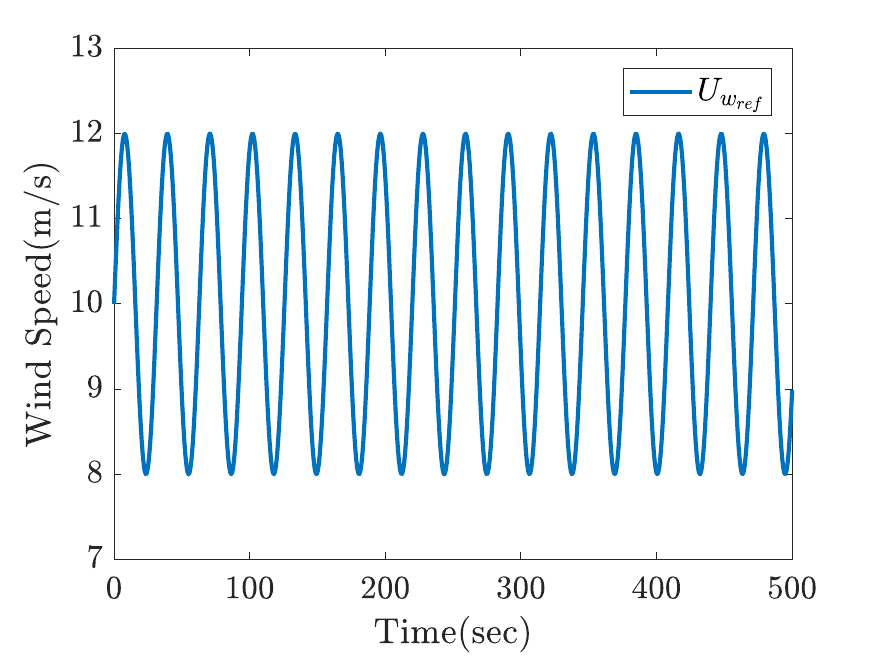}
	\caption{ The wind speed reference during the second stage.}
	\label{fig: train2 wind ref}
\end{figure}

The subsequent training stage is designed to present a richer experience to obtain a control strategy capable of handling wind speeds that exhibit constant variations. In order to accomplish this, the wind, ${U_{w}}_{ref}$, is taken as a sine wave, which can be observed in Figure \ref{fig: train2 wind ref}, and defined as follows:

\begin{equation}
{U_{w}}_{ref} = 10 + 2\sin(0.2t)
\end{equation}

\begin{figure}[t!]
	\centering
	\begin{subfigure}[b]{0.48\linewidth}
		\includegraphics[width=\linewidth]{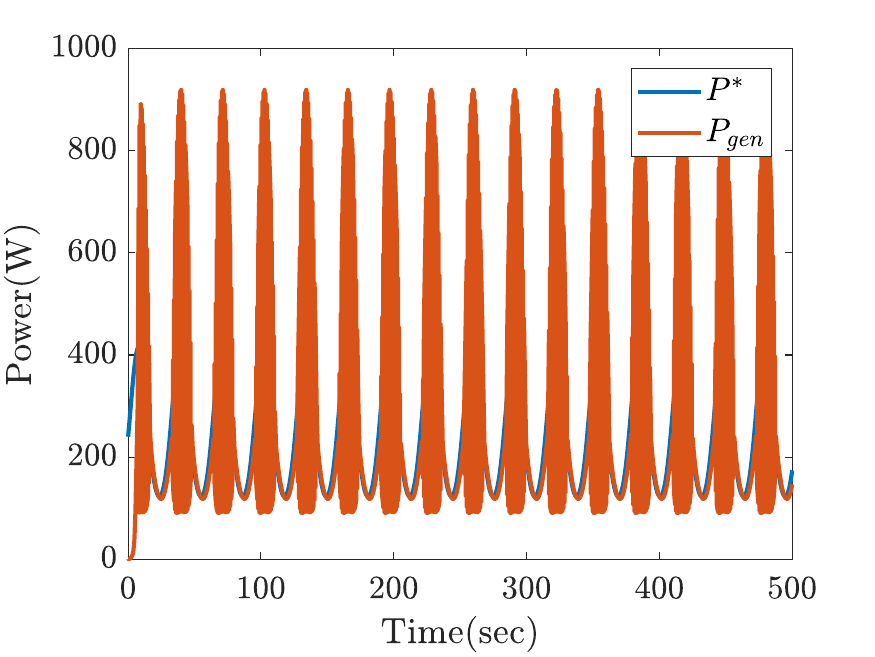}
		\caption{Obtained mechanical power}
		\label{subfig: train2_a}
	\end{subfigure}
	\begin{subfigure}[b]{0.48\linewidth}
		\includegraphics[width=\linewidth]{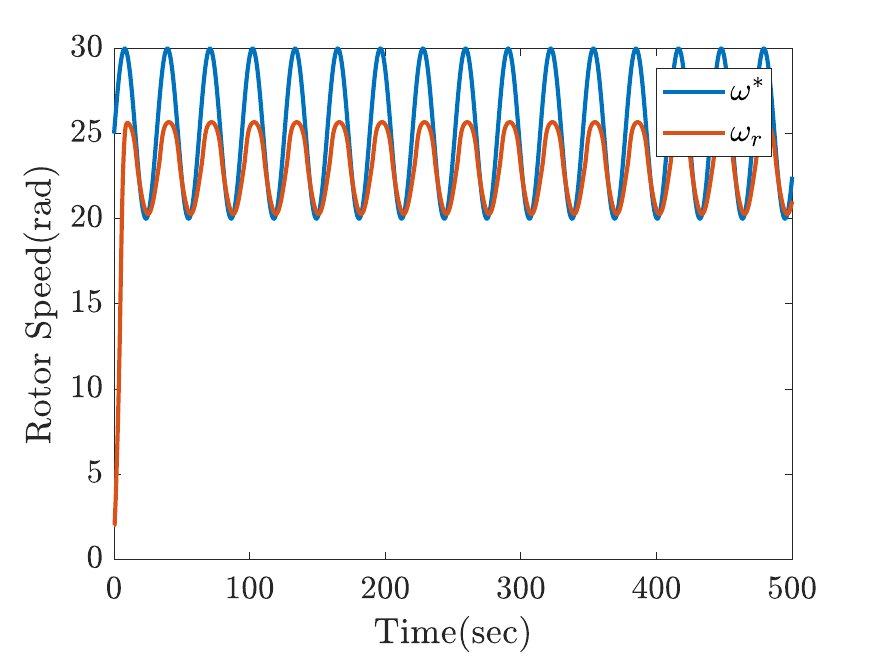}
		\caption{Rotor's angular velocity}
		\label{subfig: train2_b}
	\end{subfigure}
	\begin{subfigure}[b]{0.48\linewidth}
		\includegraphics[width=\linewidth]{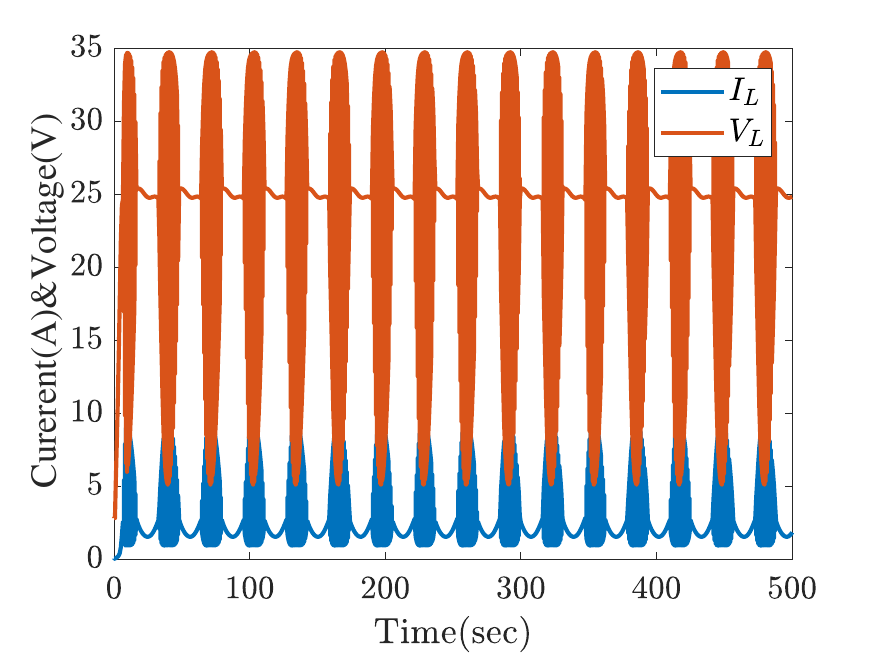}
		\caption{Measure current and voltage of the load}
		\label{subfig: train2_c}
	\end{subfigure}
	\begin{subfigure}[b]{0.48\linewidth}
		\includegraphics[width=\linewidth]{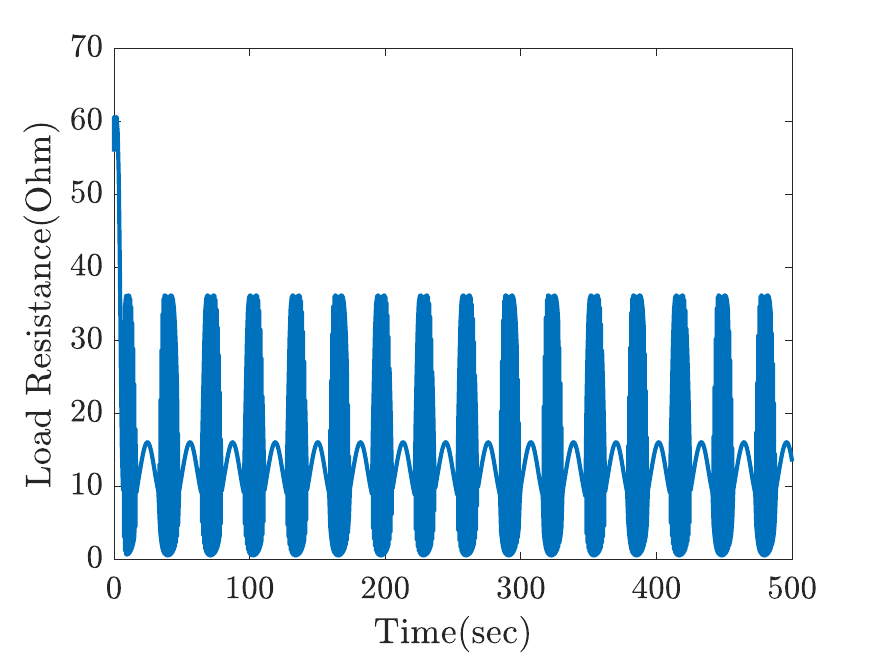}
		\caption{Resistance of the load}
		\label{subfig: train2_d}
	\end{subfigure}
	\caption{Time response results for $P$, $\omega_{r}$, $V_L$, $I_L$ and $R_L$ by RL-MCMC at the start of training's second phase using $\theta_{S1}$.}
	\label{fig: 2_stage_initial_sim}
\end{figure}

For the second phase of training, we use parameter set $\theta_{S1}$ from the result of the first training phase as our initial parameter set. Figure \ref{fig: 2_stage_initial_sim} shows generator's performance with $\theta_{S1}$ under a sine wave wind pattern (before second stage learning), and will be used to compare the proposed methodology's performance using $\theta_{S2}$, which will be obtained after second stage training is completed. 

From Figure \ref{subfig: train2_d}, 
load resistance can be seen to experience noisy peaks leading to the same peak profiles for both generated power and load voltage. 
First stage training has not provided any opportunity to make use of the derivative terms of the RBFNN inputs in the presence of constantly fluctuating winds.

\begin{figure}[t!]
	\label{fig: fig13}
	\centering
	\begin{subfigure}[b]{0.45\linewidth}
		\includegraphics[width=\linewidth]{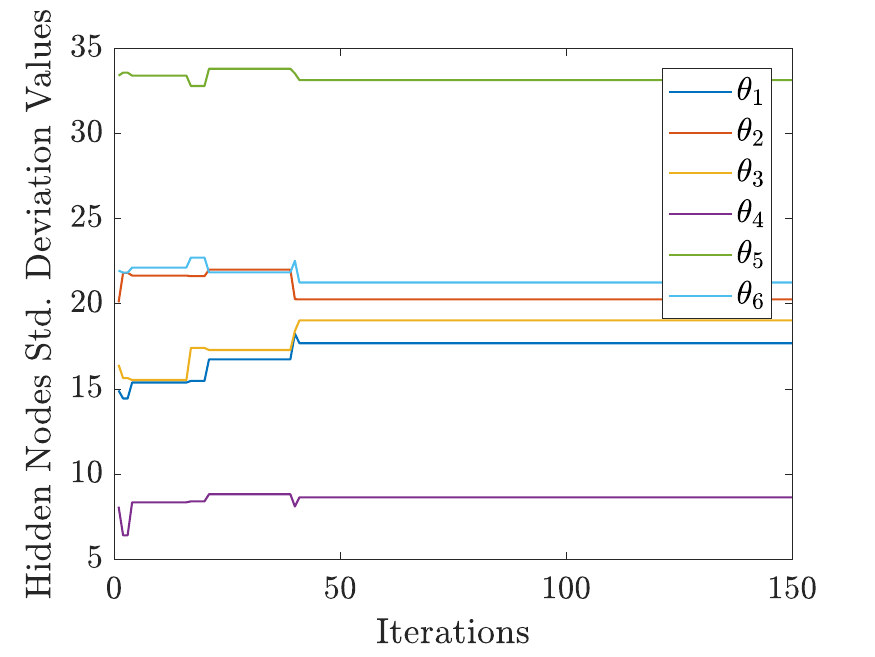}
		\caption{~Learned standard deviations}
		\label{subfig: train2_theta_a}
	\end{subfigure}
	\begin{subfigure}[b]{0.45\linewidth}
		\includegraphics[width=\linewidth]{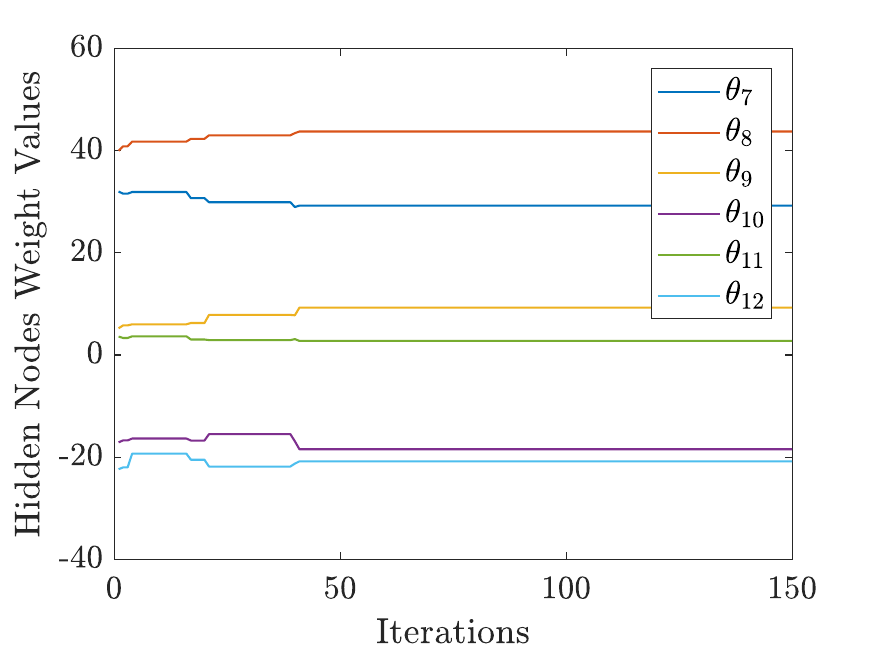}
		\caption{~Learned weights}
		\label{subfig: train2_theta_b}
	\end{subfigure}
	\caption{Trace plots of the parameters using RL-MCMC during the second stage of training.}
	\label{fig: train2_output}
\end{figure}

Figures \ref{subfig: train2_theta_a} and \ref{subfig: train2_theta_b} show the trace plots for the policy parameters during second phase training. After roughly $40$ iterations, it is clear that RL-MCMC has learned the policy parameters for the sinusoidal reference. 

\begin{figure}[t!]
	\centering
	\begin{subfigure}[b]{0.48\linewidth}
		\includegraphics[width=\linewidth]{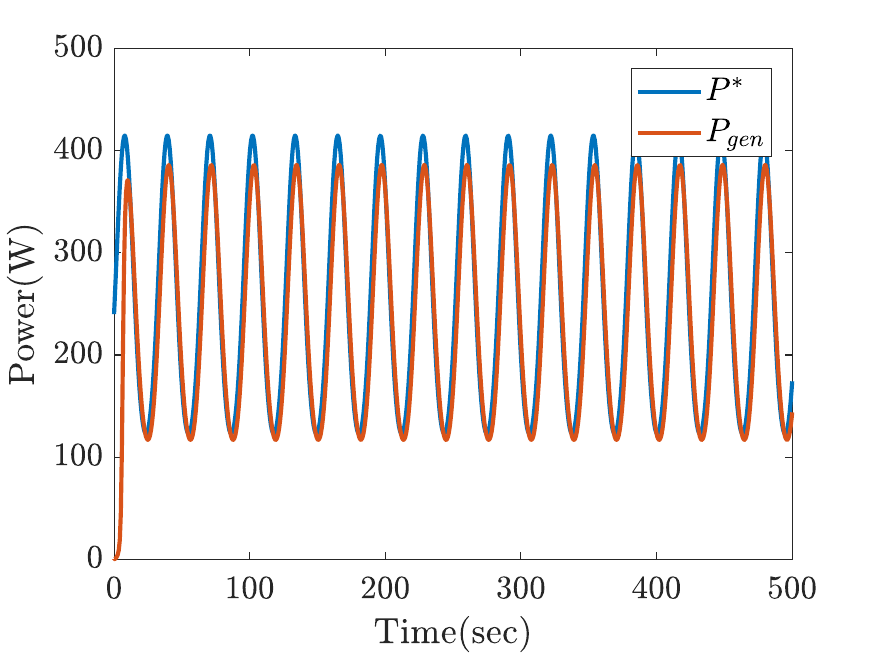}
		\caption{Obtained mechanical power}
	\end{subfigure}
	\begin{subfigure}[b]{0.48\linewidth}
		\includegraphics[width=\linewidth]{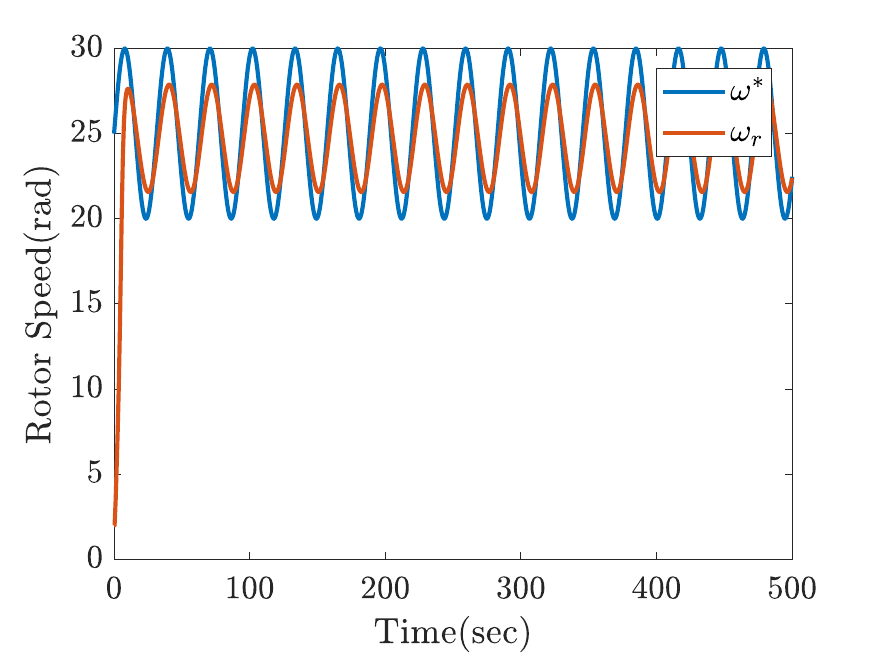}
		\caption{Rotor's angular velocity}
	\end{subfigure}
	\begin{subfigure}[b]{0.48\linewidth}
		\includegraphics[width=\linewidth]{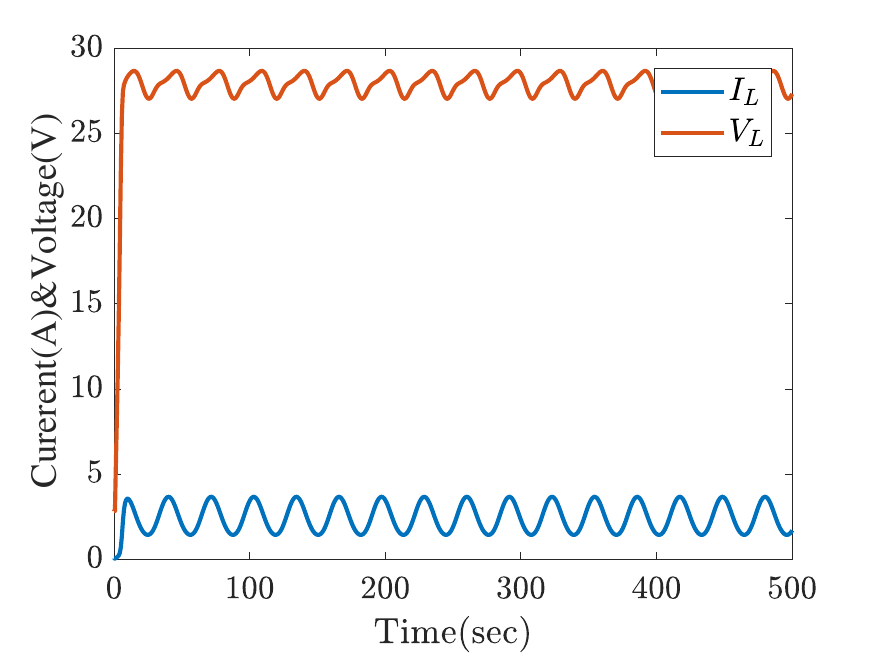}
		\caption{Measure current and voltage load}
	\end{subfigure}
	\begin{subfigure}[b]{0.48\linewidth}
		\includegraphics[width=\linewidth]{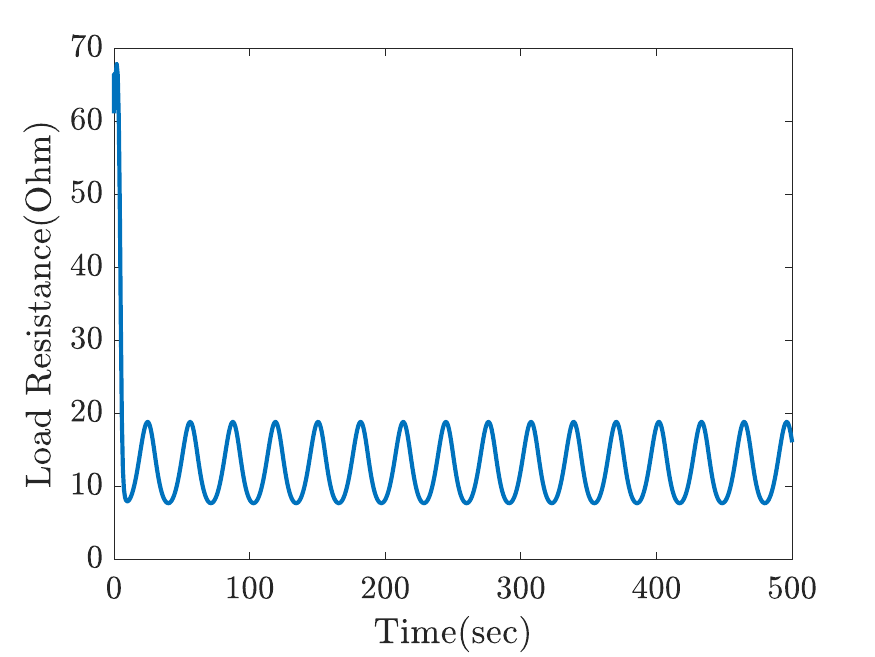}
		\caption{Resistance of the load}
	\end{subfigure}
	\caption{Time response results by RL-MCMC after $2^{nd}$ stage training, using $\theta_{S2}$ parameters}
	\label{fig: train2_output elec}
\end{figure}
Figure \ref{fig: train2_output elec} displays the results after second stage training using $\theta_{S2}$ parameters. The rotor speed and output power track the theoretically nominal power of the system, in comparison to Figure \ref{fig: 2_stage_initial_sim}. As a result, we may conclude that the system's effectiveness has been enhanced with the incorporation of training's second phase. The learned corresponding $w$ and $\sigma$ parameters for both phases are given in Table \ref{tab: training_params}.
\begin{table}
	\begin{center}
		\caption{Learned parameters for both $1^{st}$ and $2^{nd}$ training phases}
		\label{tab: training_params}
		\begin{tabular}{|c|c||c|c|} 
			\hline
			\hline
			$\boldsymbol{w_{S1}}$ & $\boldsymbol{\sigma_{S1}}$ & $\boldsymbol{w_{S2}}$ & $\boldsymbol{\sigma_{S2}}$\\
			\hline
			\hline
			$32$  & $14.9$ & $29.2$ & $17.8$\\
			$40$  & $20.1$ & $43.7$ & $20.3$\\
			$5.2$  & $16.4$ & $9.3$ & $19$ \\
			$-17.1$  & $8.1$ & $-18.5$ & $8.6$ \\
			$3.6$  & $33.4$ & $2.7$ & $33.3$ \\
			$-22.3$  & $22$ & $-20.8$ & $21.2$ \\
			\hline
		\end{tabular}
	\end{center}
\end{table}

\subsection{Comparisons between RL-MCMC, MPPT and DDPG}\label{sec: mppt,DRL}
Here, we compare the proposed RL-MCMC for WECS with the widely-used MPPT algorithm with respect to control efficiency and the total capacity of the produced energy.
In this section, a comparison between the proposed RL-MCMC for WECS and the commonly used MPPT algorithm considering control efficiency and generated energy. This comparison is carried out through two different scenarios; first, RL-MCMC and MPPT are compared using a step wind ($10~m/s$) to illustrate  onset control performances, and in the second step, the performance in a realistic wind profile is compared.
Since MPPT is a greedy algorithm aiming to maximize the instantaneous power, we expect its performance to be inferior. 
For these comparisons, $\theta_{S2}$ parameters are used in MCMC, whereas,
for MPPT, the parameter set $mppt_1$, shown in Table \ref{tab:MPPT_desc}, is used. Since it has a higher search rate, $\Delta I_{ref}$, it reaches the optimal rotor speed quickly and is suitable for realistic wind profiles, although it will present larger ripple in steady state wind.
We also tested another parameter set $mppt_2$ with smaller $\Delta I_{ref}$, designed for stable winds. Its performance will be compared briefly in Table \ref{tab: Total Energy 3}.\\

\begin{table}
	\begin{center}
		\caption{Classification of the used MPPT with respect to its parameters.}
		\label{tab:MPPT_desc}
		\begin{tabular}{|c|c|c|} 
			\hline
			\hline
			\textbf{Type} & \textbf{Sampling Time}& $\boldsymbol{\Delta I_{ref}}$\\
			\hline
			\hline
			${mppt}_1$ & $0.1~\mathrm{s}$& $0.02~\mathrm{A}$\\
			${mppt}_2$ & $0.1~\mathrm{s}$& $0.01~\mathrm{A}$\\
			\hline
		\end{tabular}
	\end{center}
\end{table}

\subsubsection{RL-MCMC and MPPT Subjected to Step Wind Speed}
The purpose of this evaluation is to contrast the onset quality of the control strategies. A given step wind as a reference is a practical method of simulating the extreme variations in the speed of wind that provide a significant challenge to WECS.   Results of RL-MCMC run with an RBFNN structure (amplitude of $10~\mathrm{m/s}$ for step velocity of wind) set to $\theta_ {S2}$ is analyzed in comparison to those of ${mppt}_{1}$  provided in Table \ref{tab:MPPT_desc}. 

\begin{figure}[t!]
	\centering
	\begin{subfigure}[b]{0.48\linewidth}
		\includegraphics[width=\linewidth]{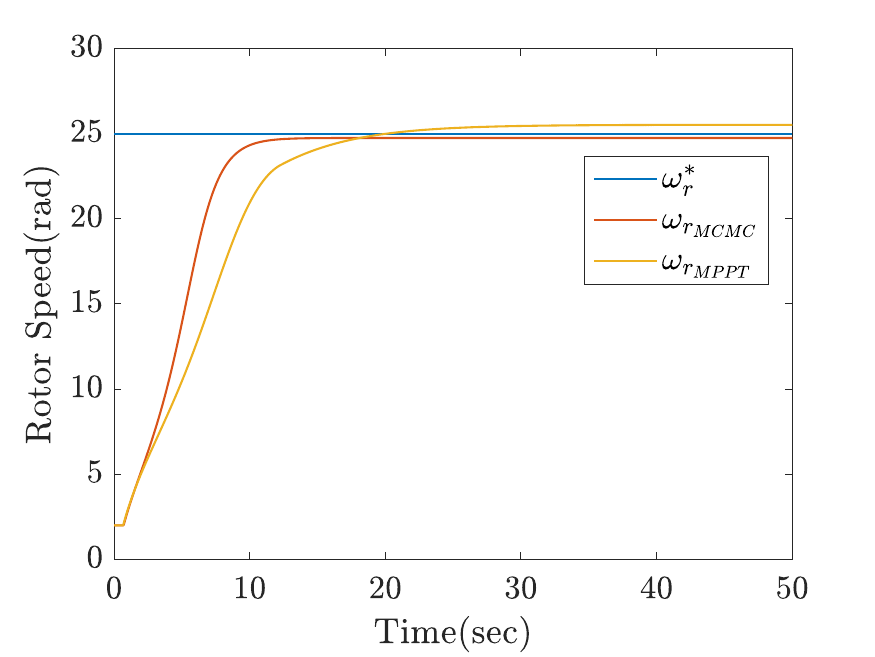}
		\caption{Rotor's nominal speed versus applied algorithms.}
		\label{subfig: rotor_comp}
	\end{subfigure}
	\begin{subfigure}[b]{0.48\linewidth}
		\includegraphics[width=\linewidth]{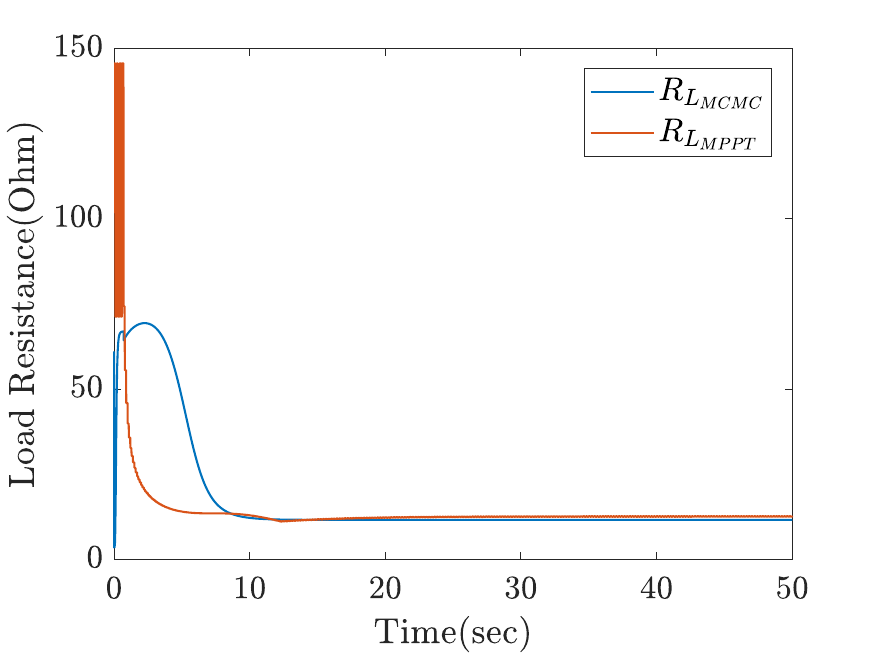}
		\caption{Resistance comparison.}
		\label{subfig: load_comp_res}
	\end{subfigure}
	\begin{subfigure}[b]{0.48\linewidth}
		\includegraphics[width=\linewidth]{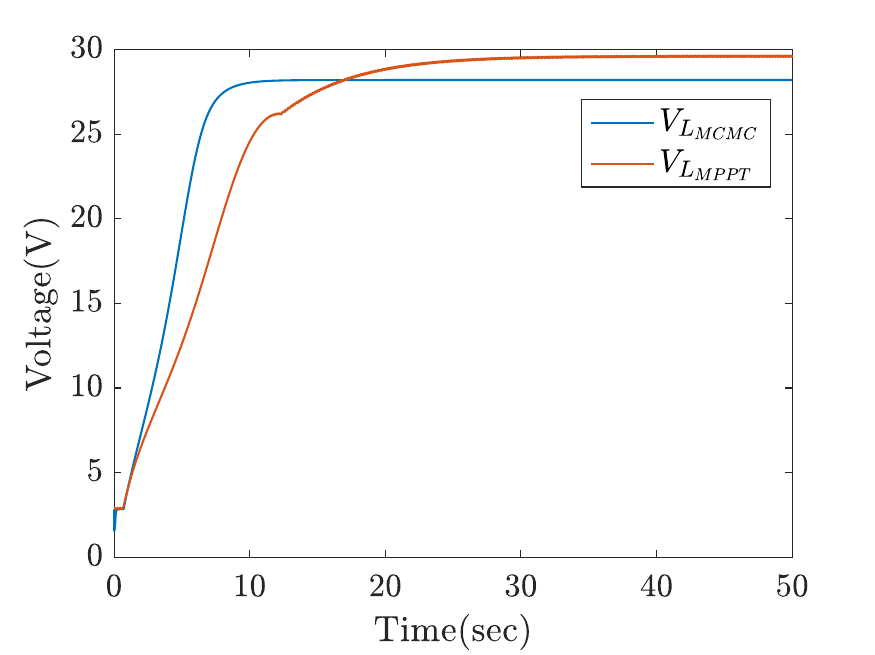}
		\caption{Voltages of the load.}
		\label{subfig: comp_voltage}
	\end{subfigure}
	\begin{subfigure}[b]{0.48\linewidth}
		\includegraphics[width=\linewidth]{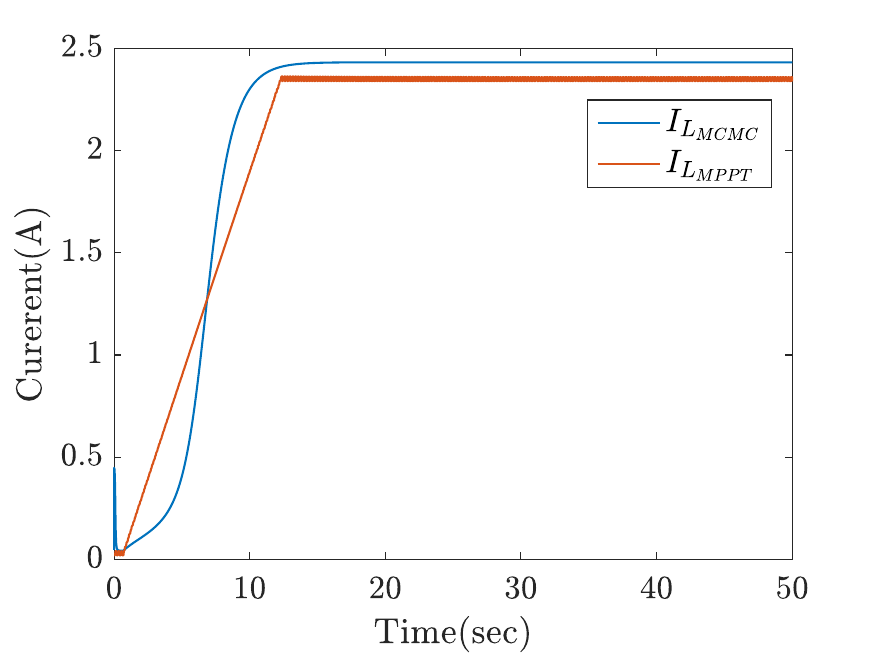}
		\caption{Currents of the load}
		\label{subfig: curr_comp}
	\end{subfigure}
	\caption{Time response comparisons between MPPT ($mppt_1$) and RL-MCMC using $\theta_{S2}$.}
	\label{fig: comp step sim 1}
\end{figure}

The simulation results of $\omega_{r}$, $R_L$, $V_L$, and $I_L$
are shown in Figure \ref{fig: comp step sim 1}. Selecting a proper $\omega_r$ is crucial to attain optimal value of $C_p$.

The value of $U_w$ is used in the equation \eqref{eq: TSR} to determine $\omega_r^{\star}$ rotor's optimum speed. $\omega_r^{\star}$ relative to MPPT and proposed RL-MCMC time plots ($\omega_{r_{MPPT}}$, $\omega_{r_{MCMC}}$) is compared in Figure \ref{subfig: rotor_comp}. One can readily see that $\omega_{r_{MCMC}}$ is more in line with $\omega_r^{\star}$ than $\omega_{r_{MPPT}}$. Figure \ref{subfig: comp_voltage} shows that $V_{L}$ is proportional to $\omega_{r}$, therefore our findings are in agreement with that. By keeping the load low at first, as demonstrated in Figure \ref{subfig: curr_comp}, RL-MCMC is able to raise $I_{L}$ smoothly and rapidly, making it the clear winner amongst the two control mechanisms.
This results in a faster convergence of MCMC to the optimal $\omega_r^{\star}$ and leads to a more efficient energy output, which was the original goal of this study. 

\begin{figure}[t!]
	\centering
	\begin{subfigure}[b]{0.48\linewidth}
		\includegraphics[width=\linewidth]{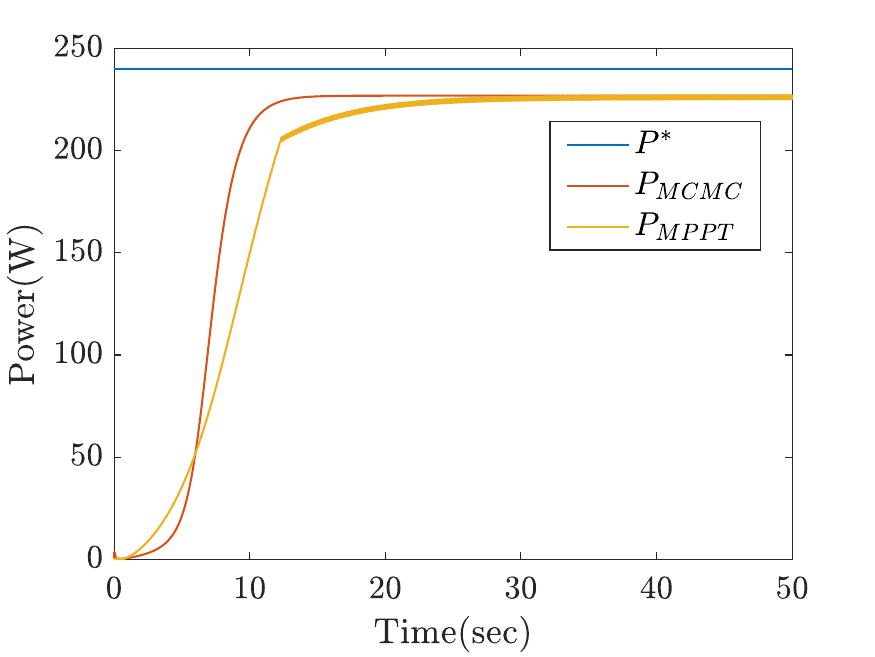}
		\caption{Obtained mechanical power.}
		\label{subfig: power_comp}
	\end{subfigure}
	\begin{subfigure}[b]{0.48\linewidth}
		\includegraphics[width=\linewidth]{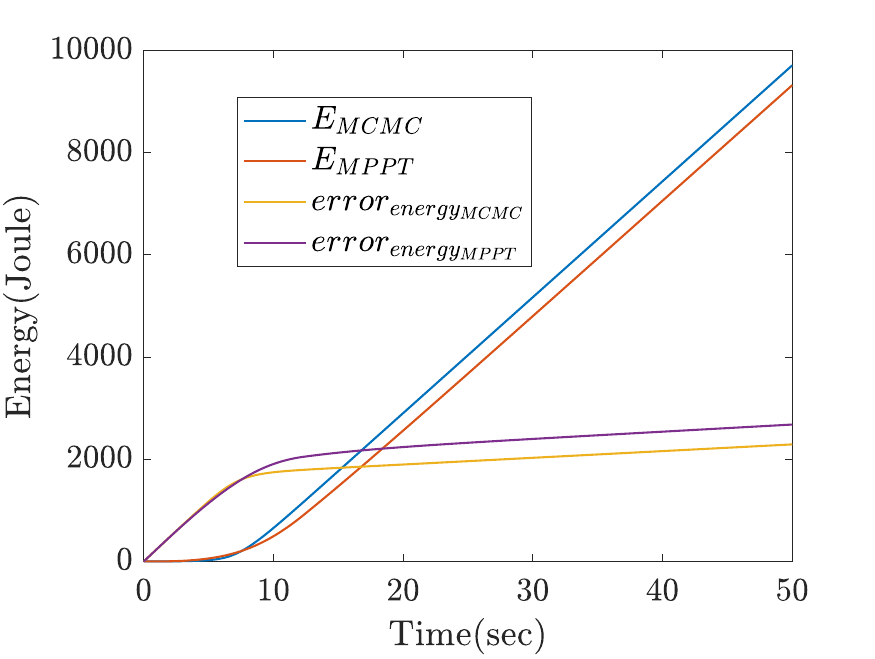}
		\caption{Output energy and energy error.}
		\label{subfig: error_energy_comp}
	\end{subfigure}
	\caption{The comparisons for the output power and energy as a result of a step wind (10~m/s).}
	\label{fig: comp step sim 2}
\end{figure}
The generated power for RL-MCMC and MPPT are demonstrated in Figure \ref{subfig: power_comp}. In steady state, they exhibit a similar performance although the ripple of MPPT output power in steady state is a disadvantage. However, the transient behaviors of these two controllers are significantly different. The total energy and energy error plots shown in Figure \ref{subfig: error_energy_comp} clearly shows the difference. MCMC performs better during the transient period and the accumulated extra energy can be seen in the $E_{MCMC}$ plot. It is expected that for rapidly and continuously changing wind profiles, the difference would be even more significant. 
  
\subsubsection{RL-MCMC and MPPT Subjected to Realistic Wind Speed}\label{subsection: real wind sim}
We evaluate the two approaches using realistic simulated wind conditions. The wind has been generated in MATLAB using the Aerospace Toolbox, where it is characterized as the combination of a variable speed with noise. For generating the wind, we have used the “Dryden Wind Turbulence Model (Continuous +q +r)” block diagram in MATLAB Aerospace Toolbox which it is used for generating a sequence of wind speeds and gusts with stationary and Gaussian distributed densities at particular altitudes. Here $q(t)$ and $r(t)$ are the inputs as longitudinal and lateral motion when used in the aircraft turbulence modeling. For the Dryden Wind Turbulence Model the parameter values that we have used are as follows:
$h = ~970 (m)$, $V = 9~(m/s)$, $DCM =$ 3 by 3 Identity Matrix, Specification as MIL-F-7885C, wind speed at low altitude intensity = 15, Wind direction = 0, Probability of exceedance of high-altitude intensity = $10^{-2}$ light, Scale length at medium/high altitudes = 553.4 (representing Dryden Turbulence), Wingspan = 10, Band limited noise sample time = 5. The output wind speed is then scaled with a gain factor of 0.7 and finally a bias of 8 is added to it.

\begin{figure}
	\centering
	\includegraphics[scale=0.6]{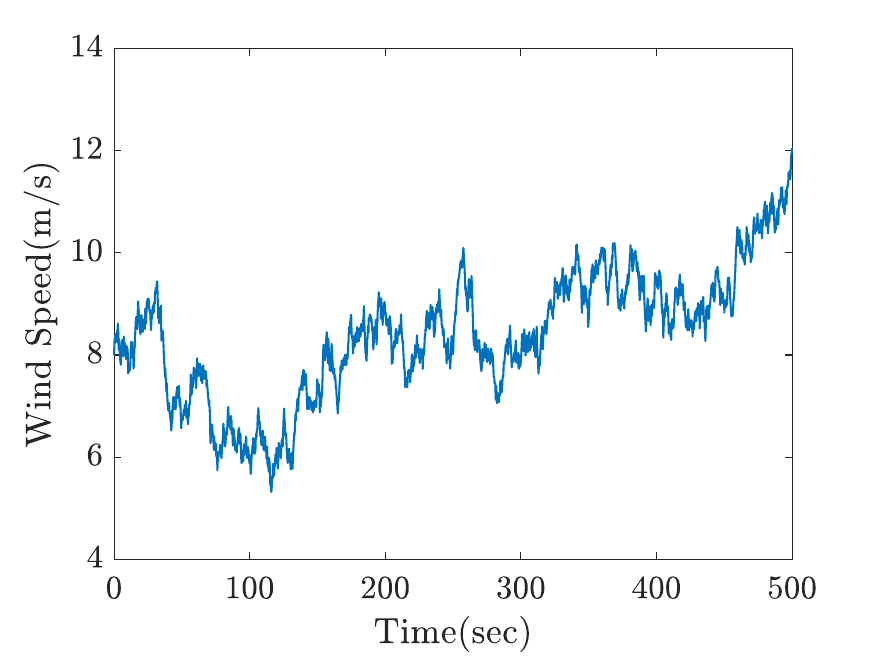}
	\caption{A sample simulated real wind speed $U_w$ using \emph{Aerospace} toolbox of MATLAB.}
	\label{fig: comp wind ref}
\end{figure}

\begin{figure}[t!]
	\centering
	\begin{subfigure}[b]{0.48\linewidth}
		\includegraphics[width=\linewidth]{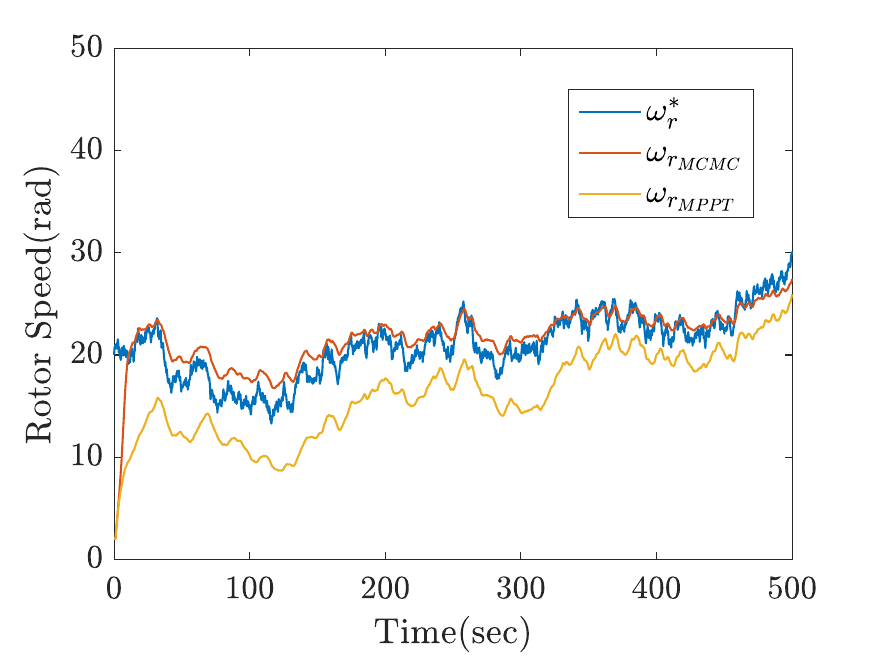}
		\caption{Rotor Speed}
		\label{fig: comp wr}
	\end{subfigure}
	\begin{subfigure}[b]{0.48\linewidth}
		\includegraphics[width=\linewidth]{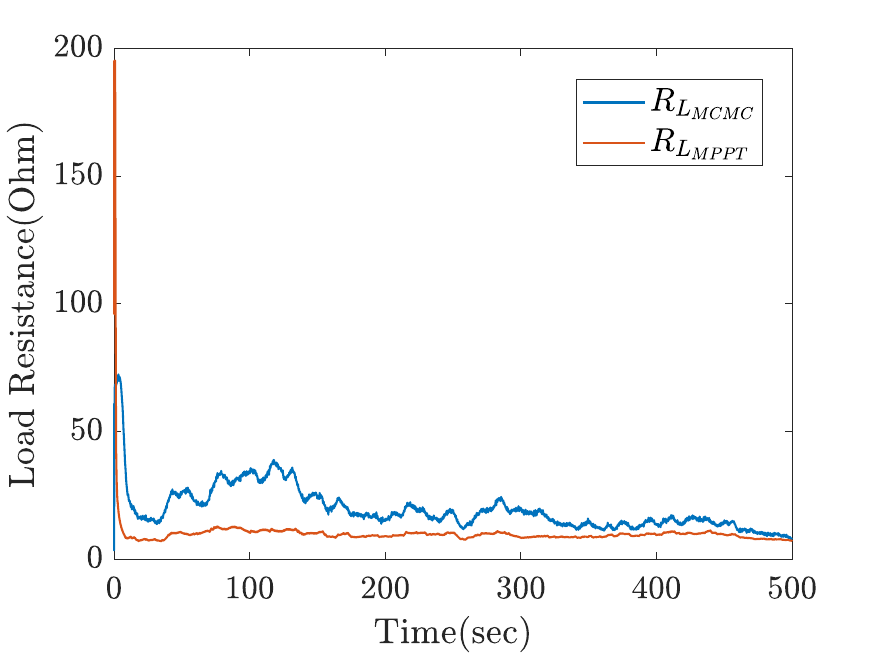}
		\caption{Load Resistance}
		\label{fig: comp rl}
	\end{subfigure}
	\caption{RL-MCMC and MPPT comparisons for the rotor speed and load resistance using a real simulated wind.}
	\label{fig: comp_rot_rezz}
\end{figure}

Figure \ref{fig: comp wind ref} displays a sample reference wind profile $U_w$ created from a given wind source. Convergence to the correct value of $\omega_r$ is the most important factor in determining the best value for $C_p$, as was discussed before. The differences in the responses between $\omega_{r_{MPPT}}$ and $\omega_{r_{MCMC}}$, are depicted in Figure \ref{fig: comp wr}. According to this figure, $\omega_{r_{MCMC}}$ is more closely maintained at $\omega^{\star}$ whereas, MPPT could not manage to generate a speed close to optimum. 

Figure \ref{fig: comp rl} shows the response of controlled resistance of the load for the RL-MCMC and MPPT where as can be seen the former one is more responsive to variations in the speed of wind.

\begin{figure}[t!]
	\centering
	\begin{subfigure}[b]{0.48\linewidth}
		\includegraphics[width=\linewidth]{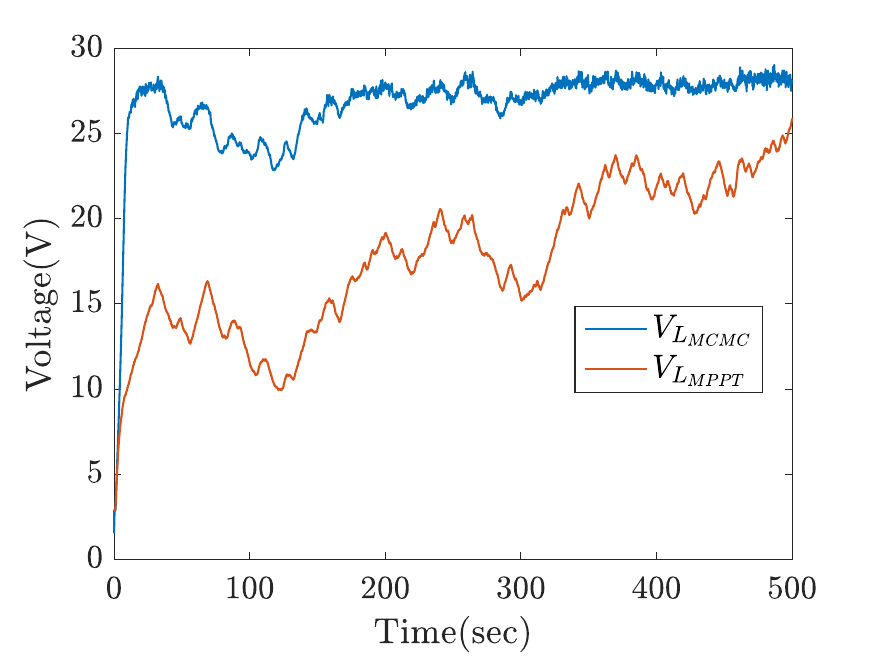}
		\caption{Load voltage}
	\end{subfigure}
	\begin{subfigure}[b]{0.48\linewidth}
		\includegraphics[width=\linewidth]{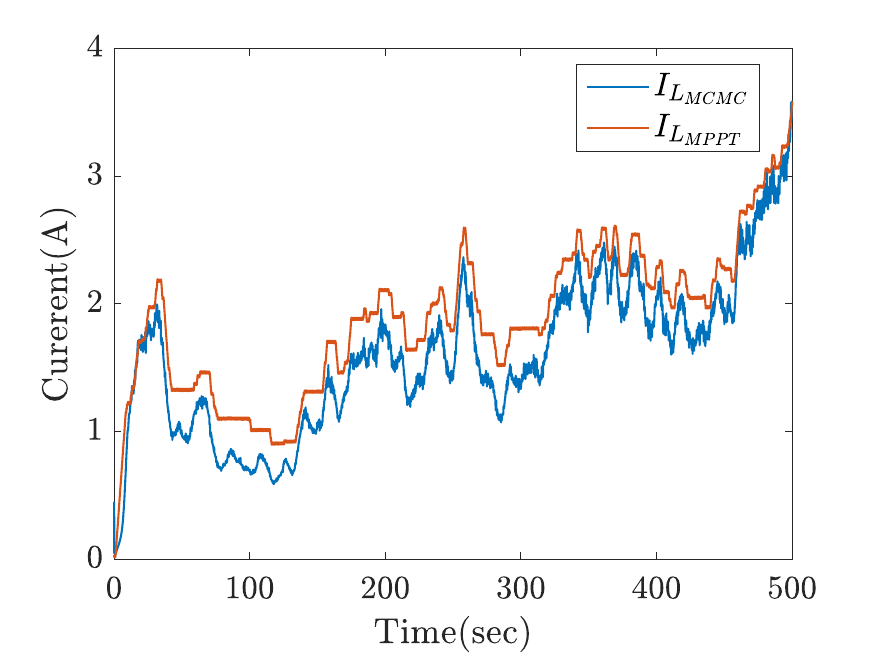}
		\caption{Load current}
	\end{subfigure}
	\caption{Comparisons between the current and voltages of the load using a real simulated wind.}
	\label{fig: comp vl_il}
\end{figure}

\begin{figure}[t!]
	\centering
	\includegraphics[scale=0.6]{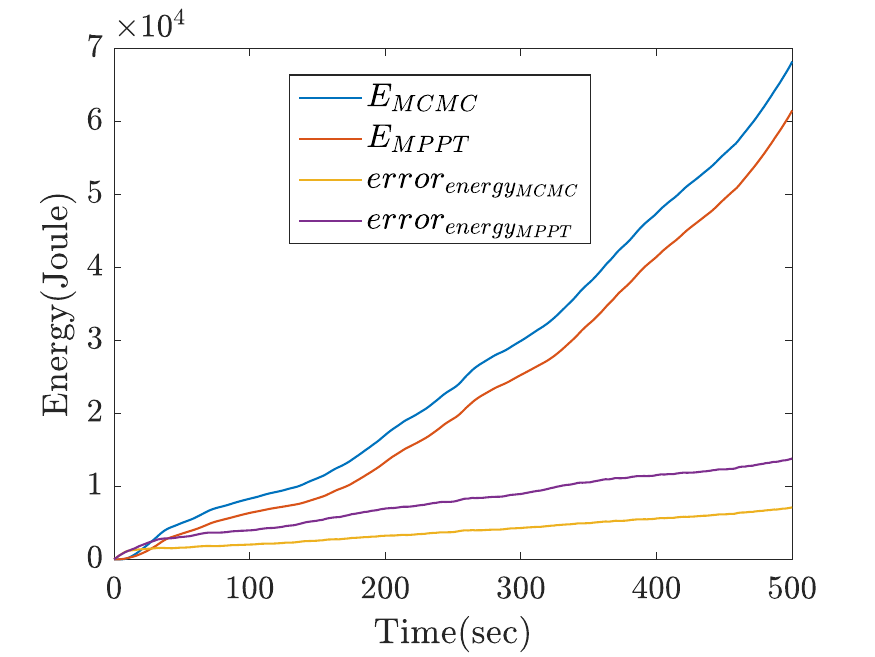}
	\caption{RL-MCMC and MPPT comparisons for the generated energy with a real simulated wind.}
	\label{fig: comp egen}
\end{figure}

The resulting response of the current and voltage in the load also are shown in Figure \ref{fig: comp vl_il}. In Figure \ref{fig: comp wr}, the wind speed does not have a definitive trend. We can see that $\omega_{r_{MCMC}}$ and $V_{L_{MCMC}}$ follow the wind speed closely, and thus they can keep a good $C_p$ value (in synchronous generators, output voltage is generally proportional to rotor speed omitting inductance and resistance effects). However, it can be seen that $V_{L_{MPPT}}$ has a definitive increasing trend which suggests that it cannot keep a good $C_p$.
\begin{figure}
	\centering
	\includegraphics[scale=0.6,]{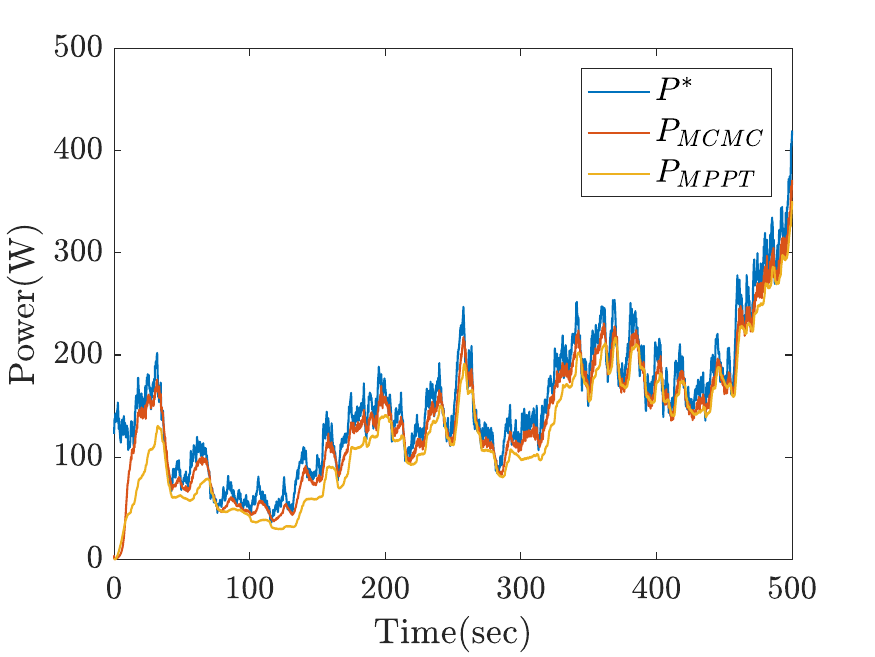}
	\caption{RL-MCMC and MPPT comparisons for the obtained mechanical power with a real simulated wind.}
	\label{fig: comp pgen}
\end{figure} 
It is evident from Figure \ref{fig: comp pgen} that the RL-MCMC's generated power can track its nominal value better than its counterpart. Figure \ref{fig: comp egen} displays the cumulative energy generated, showing that RL-MCMC can provide more energy and keeps rising above that of MPPT. 
When it comes to the error comparison, MPPT has a greater error between the ideal and actual output energy. A numerical comparison of the total wind energy to the output energy of the proposed method and MPPT can be seen in Table \ref{tab: Energy_efficient}.

\begin{table}[t!]
	\begin{center}
		\caption{Energy efficiency comparison.}
\label{tab: Energy_efficient}
\begin{tabular}{|c|c|c|} 
	\hline
	\hline
	& \textbf{Energy} & \textbf{Efficiency}\\
	\hline
	\hline
{\bf Wind Input} & $7739$ & $-$ \\
{\bf MCMC} & $6888$ & $89\%$ \\
{\bf MPPT} & $6037$ & $78\%$ \\      	
	\hline
\end{tabular}
\end{center}
\end{table} 

The statistical consistency of the two methods were also investigated. Since MCMC approaches have a statistical background, it is expected that the results be statistically more consistent. We repeated this simulation $10$ times through different real wind waves and measured total energy's mean and standard deviation (SD) for the controllers. These simulations were performed for $3$ controllers including $mppt_2$ as well as RL-MCMC and $mppt_1$. The results are listed in Table \ref{tab: Total Energy 3}. It is clear that MPPTs have nearly four times the energy output-error of the proposed algorithm.
Furthermore, Table \ref{tab: Energy mean} shows that the MPPTs' SD is high, indicating that their performance is highly sensitive to variations in wind speed, in contrast to RL-MCMC's reliable performance in a wide range of actual wind conditions.

\begin{table}[t!]
	\begin{center}
		\caption{Results of the energy output error $e$ from the  optimal (Joules).}
		\label{tab: Total Energy 3}
		\begin{tabular}{|c|c|c|c|} 
			\hline
			\hline
			\textbf{Number of tests} & \textbf{RL-MCMC's Error}& \boldmath{$mppt_1$}{\bf ~Error}& \boldmath{$mppt_2$}{\bf ~Error}\\
			\hline
			\hline
			1&8117.7& 19844.5& 12511.7\\
			2& 8254.7&9933.5&  32410.5\\
			3&8671.9& 9654.2& 11929.7\\
			4& 5917.6& 35108.9&  8024.1\\
			5& 7726.5& 33325.8&  38513.7\\
			6& 7695.6& 31365.4&  45678.7\\
			7& 7757.4& 45690.7&  8003.5\\
			8& 7787.8& 57123.1&  7802.2\\
			9& 8114.1& 22900.9&  13489.9\\
			10&  8117.4& 17834.3&  10523.4\\
			\hline
		\end{tabular}
	\end{center}
\end{table}

\begin{table}[t!]
	\begin{center}
		\caption{Mean and standard deviation of the energy output-errors (Joules).}
		\label{tab: Energy mean}
		\begin{tabular}{|c|c|c|} 
			\hline
			\hline
			\textbf{Controller}	& \textbf{Mean}& \textbf{SD}\\
			\hline
			\hline
			RL-MCMC  & 7816&732\\
			$mppt_1$& 28278& 15309\\
			$mppt_2$&23889& 27411\\
			\hline
		\end{tabular}
	\end{center}
\end{table}

\subsubsection{Comparison of RL-MCMC with DDPG Algorithm}\label{subsec: DDPG_RLMCMC}
\begin{figure}[t!]
	\centering
	\includegraphics[scale=0.5]{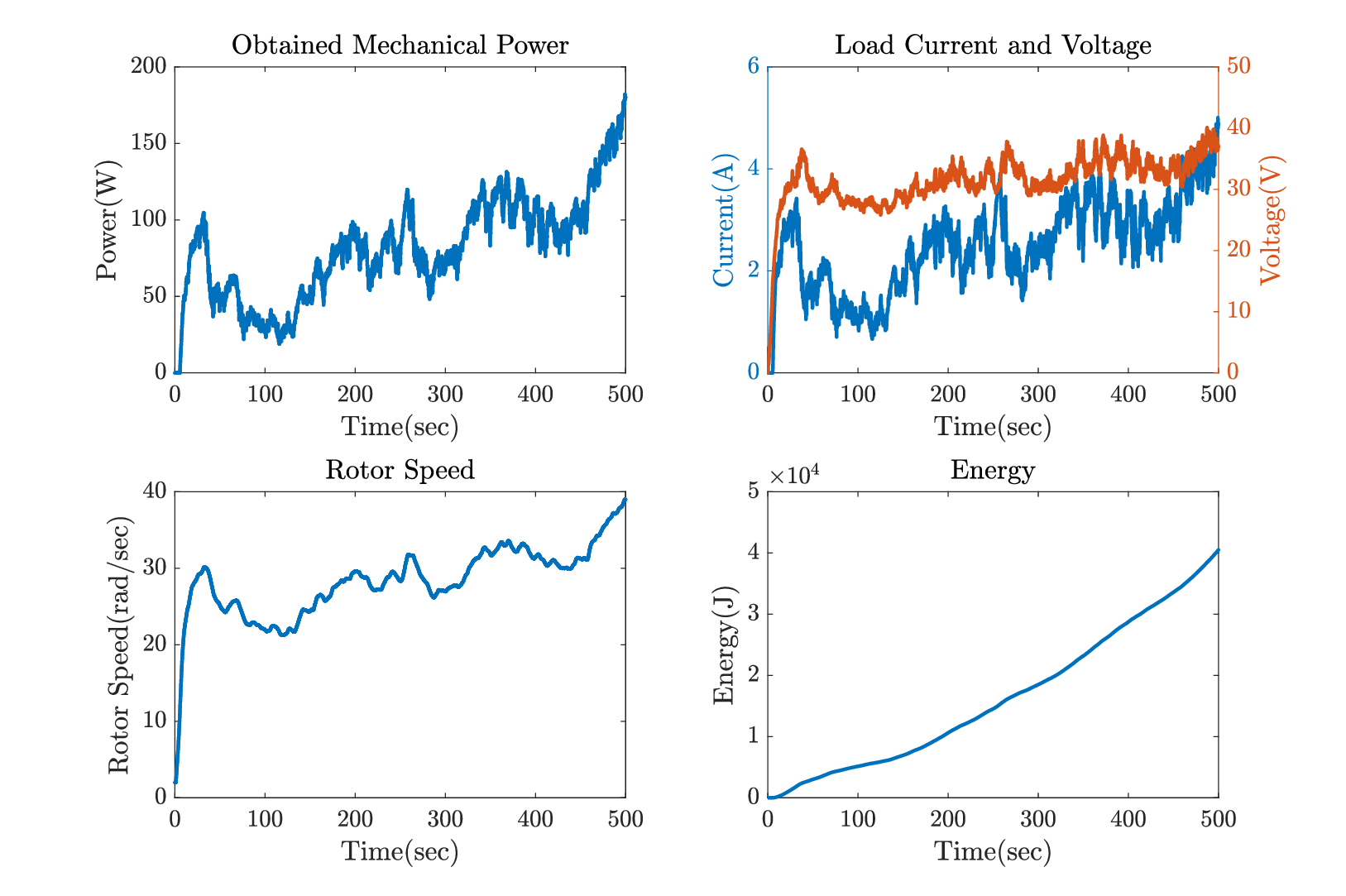}
	\caption{Time response under a real simulated wind using DDPG.}
	\label{fig: ddpg_real}
\end{figure}

A model-free and off-policy deep RL algorithm called as DDPG \citep{lillicrap2015ddpg}, which is suitable for continuous state-action spaces, is used to compare the efficiency of the proposed RL-MCMC algorithm. The action policy and value function networks are estimated by the DDPG agent using a deep neural network as a function approximator. We consider our comparison based on the real wind speed data given in section \ref{subsection: real wind sim}. The parameters of the DDPG are given in Table \ref{tab: DDPG_params}. Comparing Figures \ref{fig: comp pgen} and \ref{fig: ddpg_real} in terms of the obtained mechanical power, it is observed that the proposed RL-MCMC outperforms that of DDPG. When it comes to generated energy, as can be seen from Figure \ref{fig: comp egen}, compared to DDPG given at the bottom right of Figure \ref{fig: ddpg_real}, RL-MCMC can produce higher energy than its DDPG counterpart.

\begin{table}[t!]
	\begin{center}
		\caption{DDPG parameter values.}
		\label{tab: DDPG_params}
		\begin{tabular}{|c|c|} 
			\hline
			\hline
			\textbf{Parameter} & \textbf{Value}\\
			\hline
			\hline
			learning rate (actor) & $0.001$\\
	      	learning rate (critic) & $0.0001$\\
	      	discount factor & $0.99$\\
	      	variance decay & $0.0001$\\
	      	noise variance & $0.1$\\
	      	episodes & $300$\\
	      	mini batch size & $64$\\
	      	experience buffer size & $10^4$\\      	
			\hline
		\end{tabular}
	\end{center}
\end{table}

\section{Conclusion}\label{sec: conc}

Conventional controllers have inferior performance because of properties such as the lack of long term optimization or the requirement of linearization of the plant model which prevents them from performing well in changing wind profiles. Moreover, they may be geared towards maximizing instantaneous power output which does not result in maximum energy output. Output energy optimization of WECS is a complex problem with an complex objective function which precludes the use of gradient based algorithms. To tackle all of these problems and motivated by the
	exploration-exploitation dilemma, we have developed an RL-MCMC algorithm that combines the strengths of both
	the gradient-free Bayesian MCMC and RL algorithms. We use MCMC sampling method that is capable of exploring
	the high-reward regions of the policy space (benefiting from the RL algorithm). Since the policy
	space is explored in a non-contiguous manner, different regions can be visited and the probability of discovering better
	performing regions always exists. Our proposed Algorithm \ref{alg: Pseudo-marginal Metropolis-Hastings for reinforcement learning} does not converge to a single point. Instead, the policy parameters are guaranteed to follow the desired posterior distribution. The proposed algorithm employs an RBFNN as an approximator that learns to control the unknown nonlinear VAWT dynamics, responds to arbitrary wind profiles, and provides a nonlinear controller that manipulates the electrical load power reference as a control signal. 
When applied to learning VAWT's nonlinear model, we found that the suggested RL-MCMC technique achieves promising results. The system's performance was bootstrapped such that it could eventually handle actual wind shapes by first applying step and later sine waveform as the reference winds. We demonstrated that, in terms of overall energy production, the proposed approach outperforms its conventional counterpart MPPT. For the real wind, RL-MCMC showed to be $89$ percent energy-efficient, whereas this number for MPPT is $78$ percent. To illustrate the reliability of the proposed control method, in terms of energy output, we averaged the results of $10$ different simulations of several real winds. As may be seen in Table \ref{tab: Total Energy 3} however, MPPT was more erratic. 

As a future concern, we aim to incorporate a final training stage in a physical setup, where recorded real wind data is applied to the VAWT and corresponding final weights are learned. Since the learning can be performed on collected data with batch processing, it is possible to complete this final step using even a simple microcomputer. As another research direction, we aim at enhancing the applicability of the available RL-MCMC algorithm to even more dimensions in the state-space, using an adaptive method to update the covariance matrix of the prior distributions.
\subsection{Discussions}\label{subsec: discusions}
Based on the observations regarding the methodology and simulation results, we can point out some important discussions as listed below: 
\begin{itemize}
	\item {\bf Simple Load Model:} It should be noted that in real time applications of VAWT, the load structure is designed using Metal-Oxide Semiconductor Field-Effect Transistor (MOSFET), Insulated Gate Bipolar Transistor (IGBT), and low-Equivalent Series Resistance (ESR) capacitors. However, the simplified load circuit used in this study can still provide valuable insights into the behavior of the system and can help to inform the design of more complex systems. Also it is worth noting that the proposed RL-MCMC is a "model-free" algorithm and the complexity of the load model does not affect its performance.
	\item {\bf RL vs Other Networks:} RL is well-suited for optimizing the “long-term energy output” (and not the instantaneous energy) of wind turbines because it is a type of machine learning algorithm that focuses on decision-making in dynamic environments. In the case of wind turbines, the energy output is affected by various variables such as wind speed, turbine orientation, and other factors. RL can learn how to make decisions that maximize the energy output based on the current state of these variables, by adjusting the turbine's control inputs (e.g., rotor speed) and receiving a reward signal that reflects the long-term energy output. 
	Other types of neural networks, for example, such as supervised or unsupervised learning, are not as well-suited for this task because they typically do not have the ability to optimize an immeasurable quantity of the system, in this case, total energy. They cannot intrinsically calculate system parameters which can only be obtained through integration; again in this case the total energy is calculated from  instantaneous power. Reinforcement learning, on the other hand, is exactly suited for this purpose. Additionally, in the case of supervised learning, labeled training data for the long-term energy output is often not available. 
	\item {\bf Economical and Technical Costs:} Cost is a crucial factor to consider when choosing among the control methods. MPPT control is widely used and well established, which makes it a more cost-effective option. However the proposed RL-MCMC algorithm can feasibly be implemented on a product since it can use a batch learning process. This allows some interesting implementations:
	\begin{enumerate}
		\item The system can be implemented using a low-cost microprocessor system and learn continuously, since it can collect data and then slowly process the data without a computational deadline. This makes it possible to implement low cost and energy efficient turbines. It offers an easy-to-implement option with improved applicability and functionality for wind turbines, as it can be pretrained offline and fine-tuned online.
	   \item The system can leave the factory pre-tuned to a general performance, and then adapt to the local wind patterns. Therefore, it makes for a system with good performance out of the box that also adapts and improves as time goes on.
	   \item Long term mechanical degradation can be compensated for, due to continuous learning. It is also important to take into account the trade-off between the cost and resulting performance. While MPPT may be more cost-effective, the proposed RL-MCMC method offers superior performance in terms of the total generated energy. So, it may be worth considering the potential long-term benefits of the proposed control, such as increased efficiency and reduced energy costs, when making a decision.
	\end{enumerate}
    \item {\bf Generalization of the Proposed Method:} In fact the proposed RL-MCMC method does not need to be trained separately under different conditions. It automatically can learn in different environments and actually in this study to show its capabilities, we put it under different wind conditions. To put it better, itself trains and we believe this is an advantage because it can adapt itself to changing wind conditions including local wind patterns. Here we can mention some advantages of the RL-based control methods that make them a compelling alternative to model-bsaed algorithms such as MPPT:
    \begin{enumerate}
    	\item {\bf Adaptability:} RL-based methods can readily adapt themselves to changing conditions and environments in real-time, which allows for improved performance and greater efficiency compared to model-based methods that rely on fixed rules and models. This is possible through embedding some \emph{informative} prior knowledge to the learning process which can improve the performance and enable the agent to learn from less experience.
    	\item {\bf Flexibility:} RL-based methods can handle problems with high-dimensional state spaces, nonlinear dynamics, and complex objectives. Specifically, the proposed RL-MCMC algorithm is a model-free and data-driven algorithm which can be unaffected by parameter uncertainties and unmodelled dynamics.
    	\item {\bf Optimization:} RL-based methods can simultaneously optimize control policies based on multiple objectives, such as maximizing energy generation and minimizing mechanical degradation on the equipment.
    	\item {\bf Improved performance:} In some cases, RL-based methods have been shown to outperform model-based methods, particularly in complex or dynamic environments, such as our problem. Greedy algorithms such as MPPT try to extract power as soon as rotor speed starts to increase and therefore have worse energy output.
    	\item {\bf Real-world applications:} It can adapt itself to changing environments by sequentially updating its prior beliefs. Traditional control methods often rely on models of the system and the environment, and assume that the underlying dynamics are well understood and predictable. However, in real-world situations, these models may not be accurate enough to capture all of the complex interactions and uncertainties present in the environment.  RL-MCMC control, on the other hand, allows the control system to learn from experience and adjust its behavior in response to new and changing environmental conditions.    	 
    \end{enumerate}
   \item {\bf Optimal Policy vs a Non-Quadratic Cost Function for RL:} The estimator in Equation \ref{eq: J theta} is not quadratic. However, this is not a requirement for our method. Unlike gradient-based methods, our proposed method does not aim for maxima. Instead, it aims to provide values from high reward regions of the policy parameter and has the desire to exploit the actions with successful past experience (keeps the balance between exploration and exploitation). Being a model free method for the expected reward, the proposed method does not require a convenient shape, for the expected reward. On the contrary, the methodology in this paper is offered for cases where $J(\theta)$ does not lend itself to an easy optimization due to the nonlinearities of the system or the reward function. In principle, Algorithm \ref{alg: Pseudo-marginal Metropolis-Hastings for reinforcement learning} should work for any form of $J(\theta)$. We finally note that the estimator changes at every iteration. It is not to be considered as a function of $\theta$ that is subject to optimization. Rather, its value is used to calculate the acceptance ratio and then thrown away in the long run.
    \item {\bf Stability Analysis:} One of the limitations of the current study is the lack of stability analysis of the proposed controller. As a future work, using nonlinear control theorems like Lyaponov stability \citep{hosseini2018, javar19, bertino2022}, this idea can further be explored.
\end{itemize}



\hyphenpenalty=10000 

\bibliographystyle{cas-model2-names}			

\bibliography{refs}

\vskip6pt

\end{document}